\documentclass[aps,prl,preprint,amsmath,amssymb,showpacs,superscriptaddress,nofootinbib]{revtex4-1}
\usepackage{CJKutf8}
\usepackage{graphicx}% Include figure files
\usepackage{dcolumn}% Align table columns on decimal point
\usepackage{bm}% bold math
\usepackage{color}
\usepackage{float}
\usepackage{hyperref}% add hypertext capabilities
\allowdisplaybreaks[4]

\begin{document}

\title{Mass Probe of Tetrahedral Symmetry in Atomic Nuclei}
\author{F. F. Xu \begin{CJK*}{UTF8}{gbsn}（许方方）\end{CJK*}}
\affiliation{School of Physics, Nankai University, Tianjin 300071, China}
\author{P. W. Zhao \begin{CJK*}{UTF8}{gbsn}（赵鹏巍）\end{CJK*}}
\email{pwzhao@pku.edu.cn}
\affiliation{State Key Laboratory of Nuclear Physics and Technology, School of Physics, Peking University, Beijing 100871, China}

\date{\today}
\begin{abstract}
Tetrahedral symmetry has long been predicted as an exotic shape degree of freedom in atomic nuclei, yet clear experimental manifestations remain elusive.
We show that the triple binding energy difference $\delta V_{pn}^{(3)}$ can isolate a structural effect of tetrahedral symmetry in $^{80}$Zr.
Using relativistic density functional theory solved on a three-dimensional lattice without symmetry restrictions, the experimental $\delta V_{pn}^{(3)}$ values for even-even $^{80\text{--}90}$Zr isotopes are well reproduced without adjustable parameters.
While an enhancement of $\delta V_{pn}^{(3)}$ near $N\simeq Z$ is commonly attributed to proton-neutron correlations beyond the mean field, the pronounced nonmonotonic peak at $N=40$ emerges at the mean-field level only when the tetrahedral degree of freedom is included.
Constraining the tetrahedral deformation to zero removes the peak and leads to clear deviations from experiment.
The anomaly is traced to a well-localized tetrahedral minimum in $^{80}$Zr, supported by potential energy surfaces and characteristic single-particle level splittings.
Calculations restricted to quadrupole and triaxial shapes fail to reproduce the localized enhancement, indicating that the effect is not a generic proton-neutron correlation but a symmetry-selective increase of proton–neutron binding associated with tetrahedral geometry.
We therefore identify the $\delta V_{pn}^{(3)}$ anomaly in $^{80}$Zr as a structural mechanism distinct from the conventional Wigner-type enhancement and show that nuclear masses constitute a sensitive probe of tetrahedral symmetry.

\end{abstract}
\maketitle

\date{today}

Tetrahedral symmetry has been widely discussed in quantum systems such as molecules, metal clusters, and fullerenes.
Its possible emergence in atomic nuclei, leading to pyramid-like shapes with rounded corners and edges, has attracted considerable interest in nuclear physics~\cite{Hamamoto1991ZPD,Dudek2002PRL,Bark2010PRL,Jentschel2010PRL}, yet experimental evidence remains elusive.
Nuclear tetrahedral symmetry arises from the $T_{d}^{D}$ point group and corresponds to a nonaxial octupole ($Y_{32}$) deformation in the absence of quadrupole distortions~\cite{Dudek2002PRL}.

Possible nuclear tetrahedral shapes have been studied using various theoretical approaches, including macroscopic-microscopic models~\cite{Li1994PRC,Heiss1999PRC,Dudek2006PRL,Arita2014PRC,Dudek2018PRC}, nonrelativistic density functional theories (DFTs)~\cite{Schunck2004PRC,Zberecki2009PRC,Tagami2013PRC,Miyahara2018PRC,Wang2019PLB}, relativistic DFTs~\cite{Zhao2017PRC,Rong2023PLB,Xu2024PRC,Xu2024PLB}, the algebraic cluster model~\cite{Bijker2014PRL}, and lattice effective field theory~\cite{Epelbaum2014PRL}.
Despite these extensive studies, conclusive experimental evidence for nuclear tetrahedral shapes is still lacking~\cite{Ackermann1993NPA,Ntshangase2010PRC,Sumikama2011PRL,Hartley2017PRC}.
A key challenge is that most proposed observables are not unique to tetrahedral symmetry.
For instance, negative-parity rotational bands with suppressed $E2$ transitions~\cite{Bark2010PRL,Jentschel2010PRL} can be obscured by competing correlations and configuration mixing.
It is therefore desirable to identify observables that are sensitive to structural changes in the intrinsic single-particle spectrum.

Binding energy differences provide a complementary probe of nuclear structure, as they reflect correlation energies in a largely model-independent way.
In particular, the double binding energy difference $\delta V_{pn}$ measures the average proton-neutron interaction~\cite{Zhang1989PLB,Brenner1990PLB} and is strongly enhanced in nuclei with $N \simeq Z$.
Its magnitude correlates with a variety of nuclear properties, including the evolution of collective behavior~\cite{Casten1985PRL,Cakirli2006PRL}, the emergence of shell closures~\cite{Heyde1985PLB,Cakirli2005PRL}, and the Wigner energy~\cite{Isacker1995PRL,Satula1997PLB}.
Accordingly, $\delta V_{pn}$ has attracted sustained experimental~\cite{Chen2009PRL,Neidherr2009PRL,Ketelaer2011PRC,Mardor2021PRC,Wang2023PRL} and theoretical~\cite{Brenner2006PRC,Fu2010PRC,Bender2011PRC,Wu2016PRC,ZhangW2019PRC,Wang2024PRL,Cakirli2025FOP} interest.
Furthermore, its behavior has been linked to the onset of quadrupole prolate and oblate shapes~\cite{Federman1978PLB,Stoitsov2007PRL}.

Higher-order binding energy differences amplify local variations of underlying proton-neutron correlations.
In particular, the triple difference $\delta V_{pn}^{(3)}$ suppresses smooth systematic trends and highlights localized changes in the single-particle structure.
Since tetrahedral symmetry generates characteristic degeneracies in the single-particle spectrum, it can induce a localized enhancement of proton-neutron binding that becomes especially visible in $\delta V_{pn}^{(3)}$.
This effect has so far remained unexplored both experimentally and theoretically.

In this Letter, we investigate double and triple binding energy differences microscopically across the full deformation space using relativistic DFT on a three-dimensional (3D) lattice.
Unlike previous 3D DFT studies that focused on potential-energy surfaces (PESs) or exotic shapes, we compute binding energy differences.
The proposed triple binding energy difference provides a strong experimental signature of tetrahedral symmetry, while the 3D lattice relativistic DFT calculations provide the corresponding microscopic interpretation.
The relativistic framework naturally includes the spin-orbit interaction~\cite{Ring1996PPNP,Ren2020PRC} and, when implemented without symmetry restrictions, accounts for all deformation degrees of freedom~\cite{Ren2019SCI,Li2020PRC,Xu2024PRC}.
This approach has proven successful in describing nuclear exotic shapes~\cite{ZhangDD2022PRC,Ren2020NPA,Xu2024PRC,Xu2024PLB}.

We apply this framework to even-even Zr isotopes, which provide an ideal testbed because their single-particle structure around $N\approx40$ has long been predicted to favor nonaxial octupole correlations~\cite{Yamagami2001NPA}.
Our calculations reproduce experimental $\delta V_{pn}^{(3)}$ values for $^{80\text{--}90}$Zr without adjustable parameters.
A pronounced and nonmonotonic peak appears at $N=40$, and remarkably this anomaly is obtained at the mean-field level only when the tetrahedral degree of freedom is allowed; constraining it to zero eliminates the anomaly.
While $\delta V_{pn}$ enhancements are common in many $N \simeq Z$ nuclei, the $^{80}$Zr peak originates from a distinct structural mechanism associated with a well-localized tetrahedral minimum and characteristic single-particle level splittings.
These results indicate that the mass anomaly of $^{80}$Zr is not a generic Wigner-type correlation but a symmetry-selective enhancement of proton–neutron binding due to tetrahedral geometry, providing a new avenue to identify tetrahedral symmetry in nuclei.

The starting point of relativistic DFT is a universal energy density functional of nucleon densities and currents~\cite{Meng2016Book,Vretenar2005PR,Niksic2011PPNP}.
The variational principle leads to the single-nucleon Dirac equation
\begin{eqnarray}
  \label{eq:1}
   \left\{\bm{\alpha}\cdot \bm{p}  + \beta\left[m+S(\bm{r})\right]+V(\bm{r})\right\}\psi_{k}=\varepsilon_{k}\psi_{k},
\end{eqnarray}
where $m$ is the nucleon mass and $\varepsilon_{k}$ denotes the single-particle energy.
The scalar field $S(\bm{r})$ and vector field $V(\bm{r})$ are generated self-consistently from the nucleon densities of the occupied orbitals.
In the present work, the Dirac equation~\eqref{eq:1} is solved directly on a 3D lattice without imposing any symmetry, which allows a fully microscopic description of nuclear arbitrary deformations.

Pairing correlations are treated in the Bardeen–Cooper–Schrieffer (BCS) approximation using the pairing functional
\begin{eqnarray}
  \label{eq:2}E_{{\rm pair}}=-\sum_{\tau=n,p}\frac{G_{\tau}}{4}\int d^3{\bm r}\kappa_{\tau}^{*}(\bm r)\kappa_{\tau}(\bm r),
\end{eqnarray}
where $G_{\tau}$ is the pairing strength and the pairing tensor is defined as
\begin{eqnarray}
  \label{eq:3}\kappa(\bm r)=2\sum_{k>0}f_{k}u_{k}v_{k}|\psi_{k}(\bm r)|^2.
\end{eqnarray}
The smooth-cutoff factor is given by
\begin{eqnarray}
  \label{eq:4}f_{k}=\frac{\Theta(-\varepsilon_k)}{1+{\rm exp}[(\varepsilon_{k}-\lambda_{{\rm F}}-\Delta E_{\tau})/\mu_{\tau}]},
\end{eqnarray}
where $\Delta E_{\tau}=5$ MeV and $\mu_{\tau}=0.5$ MeV are adopted from Ref.~\cite{Ryssens2015CPC}.
The step function $\Theta(-\varepsilon_k)$ equals one for bound levels and zero elsewhere.

The double binding energy difference are extracted from the calculated total energies for even-even nuclei~\cite{Isacker1995PRL},
\begin{eqnarray}
  \label{eq:5}
  \delta V_{pn}(Z,N)=-\frac{1}{4}\big[E(Z,N)-E(Z,N-2)-E(Z-2,N)+E(Z-2,N-2)\big],
\end{eqnarray}
and the triple binding energy difference
\begin{eqnarray}
  \label{eq:6}\delta V_{pn}^{(3)}(Z,N)=\delta V_{pn}(Z,N)-\delta V_{pn}(Z,N+2).
\end{eqnarray}

Nuclear shapes are characterized through multipole moments calculated from the converged vector density $\rho_{v}(\bm r)$ in a fixed intrinsic frame, defined by aligning the principal axes of inertia with the coordinate axes of the 3D lattice.
Quadrupole deformations are described by the moments $a_{20}$ and $a_{22}$, or equivalently by the parameters $\beta$ and $\gamma$,
\begin{subequations}
  \begin{eqnarray}
    \label{eq:7a}a_{20}=&&\frac{4\pi}{3AR^{2}}\int d^3r\rho_{v}(\bm r)r^{2}Y_{20}=\beta\cos\gamma,\\
    \label{eq:7b}a_{22}=&&\frac{4\pi}{3AR^{2}}\int d^3r\rho_{v}(\bm r)r^{2}Y_{22}=\frac{1}{\sqrt{2}}\beta\sin\gamma.
  \end{eqnarray}
\end{subequations}
Following Ref.~\cite{Xu2024PRC}, a pure octupole shape is represented by the four independent moments $a_{3m}$ with $m=0,1,2,3$,
\begin{subequations}
  \begin{eqnarray}
    \label{eq:8a}a_{30}=&&\frac{4\pi}{3AR^{3}}\int d^3r\rho_{v}(\bm r)r^{3}Y_{30},\\
    \label{eq:8b}a_{31}=&&\frac{4\pi}{3AR^{3}}\int d^3r\rho_{v}(\bm r)r^{3}\left[-\frac{\sqrt{5}}{4}(Y_{33}-Y_{3-3})+\frac{\sqrt{3}}{4}(Y_{31}-Y_{3-1})\right],\\
    \label{eq:8c}a_{32}=&&\frac{4\pi}{3AR^{3}}\int d^3r\rho_{v}(\bm r)r^{3}\left[-\frac{i}{2}(Y_{32}-Y_{3-2})\right],\\
    \label{eq:8d}a_{33}=&&\frac{4\pi}{3AR^{3}}\int d^3r\rho_{v}(\bm r)r^{3}\left[-\frac{\sqrt{5}i}{4}(Y_{33}+Y_{3-3})-\frac{\sqrt{3}i}{4}(Y_{31}+Y_{3-1})\right].
  \end{eqnarray}
\end{subequations}
To uniquely define all possible octupole shapes without a repetition, the range of parameters is restricted to $a_{32}\geq0$ and $a_{30}\geq a_{31}\geq a_{33}\geq 0$.

In this work, the triple binding energy difference $\delta V_{pn}^{(3)}$ for even-even Zr isotopes is investigated using relativistic DFT on a 3D lattice space.
The lattice spacing is chosen as $1$ fm and the grid numbers along the $x$, $y$ and $z$ axes are $30$.
The point-coupling energy density functional PC-PK1~\cite{Zhao2010PRC} is employed.
Consistent with Refs.~\cite{Zhao2017PRC,Xu2024PRC}, the neutron and proton pairing strengths are taken as
$G_{n}= -330~{\rm MeV\,fm}^{3}$ and $G_{p}= -430~{\rm MeV\, fm}^{3}$,
which reproduce the empirical pairing gaps of Zr isotopes.

\begin{figure*}[!htbp]
  \centering
  \includegraphics[width=1.00\textwidth]{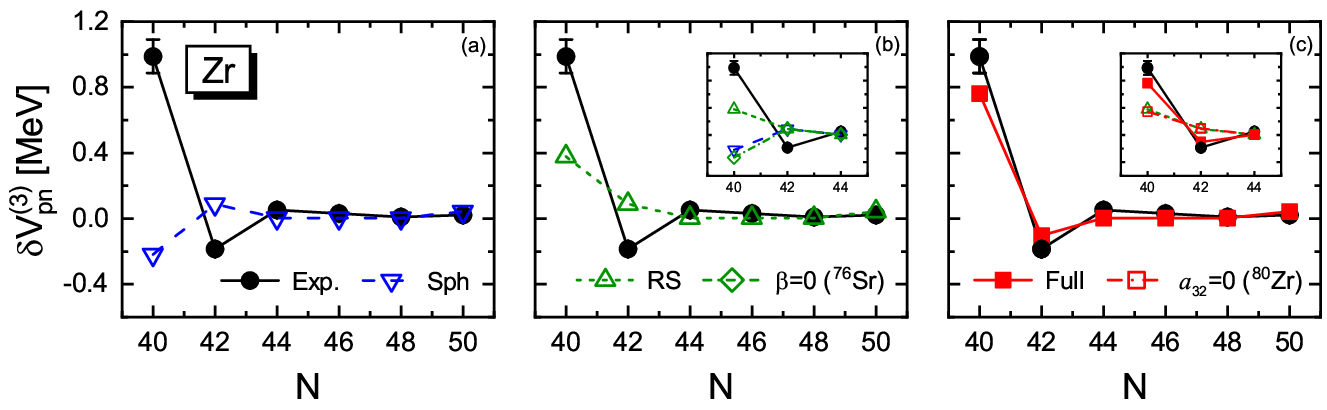}
  \caption{(Color online.) Calculated $\delta V_{pn}^{(3)}$ for even-even $^{80\text{--}90}$Zr compared with experimental data (solid circles) extracted from the AME2020 data~\cite{Wang2021CPC}, including the extrapolated values.
  The data of $^{80}$Zr is obtained from the recent high precision mass measurement~\cite{Hamaker2021Nature}.
  Results with spherical symmetry (Sph), reflection symmetry (RS), and full 3D deformation (Full) are denoted by open down triangles, open up triangles, and solid squares, respectively.
  Insets show the detailed structure near $N=42$: in panel (b), open diamonds correspond to calculations with $^{76}$Sr constrained to be spherical; in panel (c), open squares correspond to calculations with the tetrahedral deformation of $^{80}$Zr set to zero.
   }
  \label{Fig1}
\end{figure*}

According to Eqs.~\eqref{eq:5} and~\eqref{eq:6}, studying the triple binding energy difference $\delta V_{pn}^{(3)}$ for even-even Zr isotopes requires the ground-state binding energies of the neighboring even-even Zr and Sr nuclei.
Figure~\ref{Fig1} compares the calculated $\delta V_{pn}^{(3)}$ values for $^{80\text{--}90}$Zr under different symmetry restrictions with experimental values, extracted from the atomic mass evaluation (AME2020)~\cite{Wang2021CPC} and Ref.~\cite{Hamaker2021Nature} for $^{80}$Zr.
Along the isotopic chain, $\delta V_{pn}^{(3)}$ remains nearly constant at 0.01--0.05 MeV for $N=44$--50.
In contrast, from $N=40$ to $N=44$, it exhibits a pronounced, nonmonotonic anomaly: reaching 0.99 MeV for $N=40$, dropping sharply to $-0.19$ MeV at $N=42$, and  partially recovering to about 0.05 MeV at $N=44$.
This striking behavior calls for a microscopic explanation.

To explore the origin, we calculate $\delta V_{pn}^{(3)}$ under different symmetry restrictions.
As shown in Fig.~\ref{Fig1}(a), restricting all nuclei to spherical shapes reproduces the experimental values for $N=44$--50, but fails for $^{80,82}$Zr, even inverting the anomalous trend.
This indicates that additional shape degrees of freedom are essential for these nuclei, while isotopes with $N\ge44$ should remain nearly spherical.

Allowing quadrupole deformation under reflection symmetry [Fig.~\ref{Fig1}(b)] leaves $\delta V_{pn}^{(3)}$ nearly unchanged for $N=42$--50, confirming that the corresponding Zr and Sr isotopes possess nearly spherical ground states.
For $^{80}$Zr, however, $\delta V_{pn}^{(3)}$ rises from $-0.22$ to 0.38 MeV because the quadrupole deformation lowers the ground-state energy of $^{76}$Sr.
As shown in the inset, enforcing $\beta=0$ for $^{76}$Sr (open diamonds) brings the result for $^{80}$Zr back to the spherical-symmetry value.
Nevertheless, allowing quadrupole deformation still does not reproduce the observed anomalies for $^{80}$Zr and $^{82}$Zr, implying that higher-order shape degrees of freedom are required.

Removing all symmetry restrictions in the 3D lattice calculation, as shown in Fig.~\ref{Fig1}(c), reproduces the experimental $\delta V_{pn}^{(3)}$ values as well as the anomalous trend from $N=40$ to $N=44$.
Suppressing the tetrahedral degree of freedom in $^{80}$Zr by imposing $a_{32}=0$   (open squares in the inset) restores the reflection-symmetric result, unambiguously identifying tetrahedral deformation as the key origin of the anomaly.
The emergence of tetrahedral deformation lowers the binding energy and alters the single-particle level structure, thereby generating the observed $\delta V_{pn}^{(3)}$ behavior.
For $N\ge 44$, the calculated values remain nearly constant, consistent with the experimental data.
These results demonstrate that $\delta V_{pn}^{(3)}$ serves as a sensitive indicator of tetrahedral deformation and firmly identify this degree of freedom in $^{80}$Zr as the origin of the anomaly.

\begin{figure*}[!htbp]
  \centering
  \includegraphics[width=0.90\textwidth]{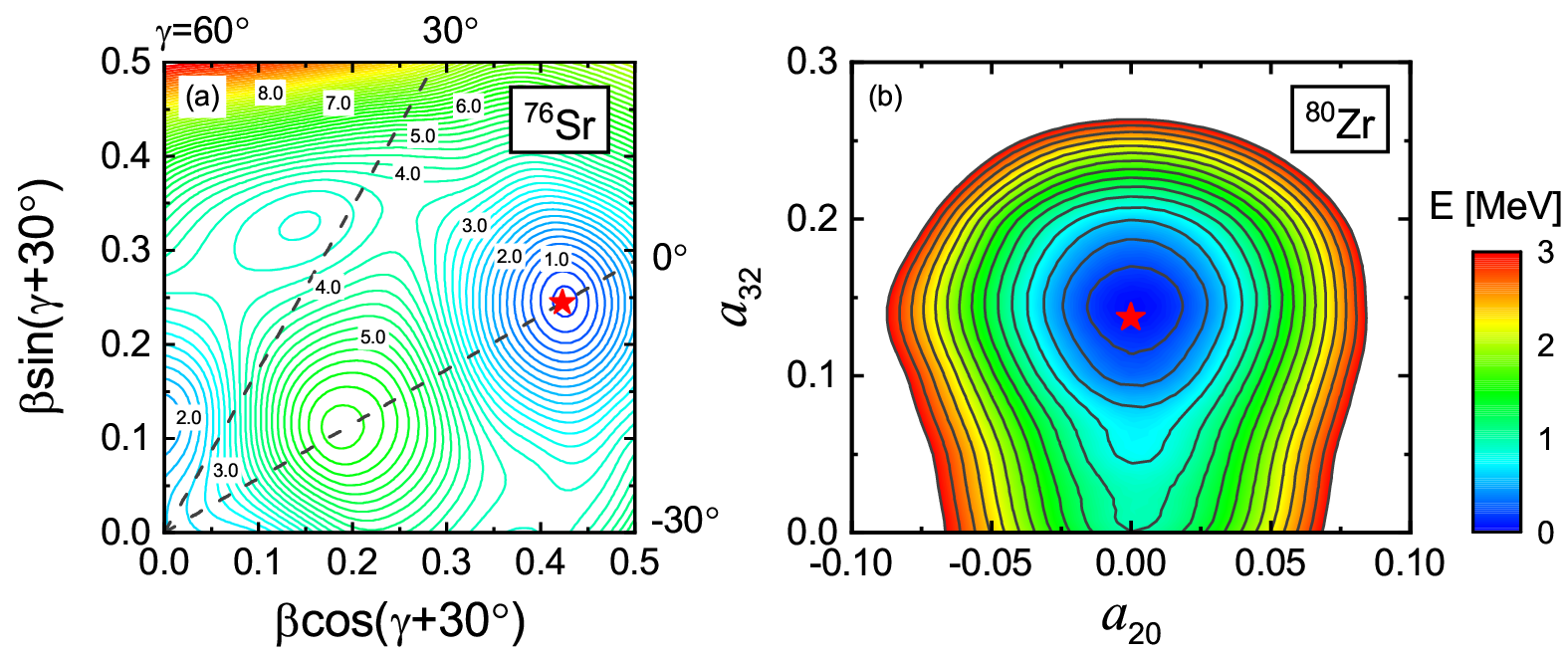}
  \caption{(Color online.) Potential energy surfaces of $^{76}$Sr in the $(\beta,\gamma)$ plane (a) and of $^{80}$Zr in the $(a_{20},a_{32})$ plane (b).
  The energy minimum is indicated by a solid star.
  Energies are normalized to the respective minimum, and the contour interval is 0.2 MeV.
  }
  \label{Fig2}
\end{figure*}

From the above discussions, it is clear that allowing quadrupole and octupole degrees of freedom lowers the ground-state energies of $^{76}$Sr and $^{80}$Zr, thereby modifying the $\delta V_{pn}^{(3)}$ values.
To quantify these energy gains and identify the equilibrium shapes, the calculated PESs of $^{76}$Sr in the $(\beta,\gamma)$ plane and $^{80}$Zr in the $(a_{20},a_{32})$ plane are shown in Fig.~\ref{Fig2}.

The PES of $^{76}$Sr exhibits a well-defined prolate minimum at $\beta=0.49$ and $\gamma=0^\circ$, with steep contours indicating a robust ground state and an energy gain of $\sim2.84$ MeV.
For $^{80}$Zr, the PES shows a well-localized tetrahedral minimum at $a_{20}=0$ and $a_{32}=0.14$, with steep curvature along $a_{20}$ and moderately bounded curvature along $a_{32}$.
The associated energy gain of $\sim0.84$ MeV fully accounts for the anomalous $\delta V_{pn}^{(3)}$ behavior near $N=42$, unambiguously identifying tetrahedral deformation as the decisive structural origin, while quadrupole effects alone fail to reproduce the anomaly.

\begin{figure*}[!htbp]
  \centering
  \includegraphics[width=0.78\textwidth]{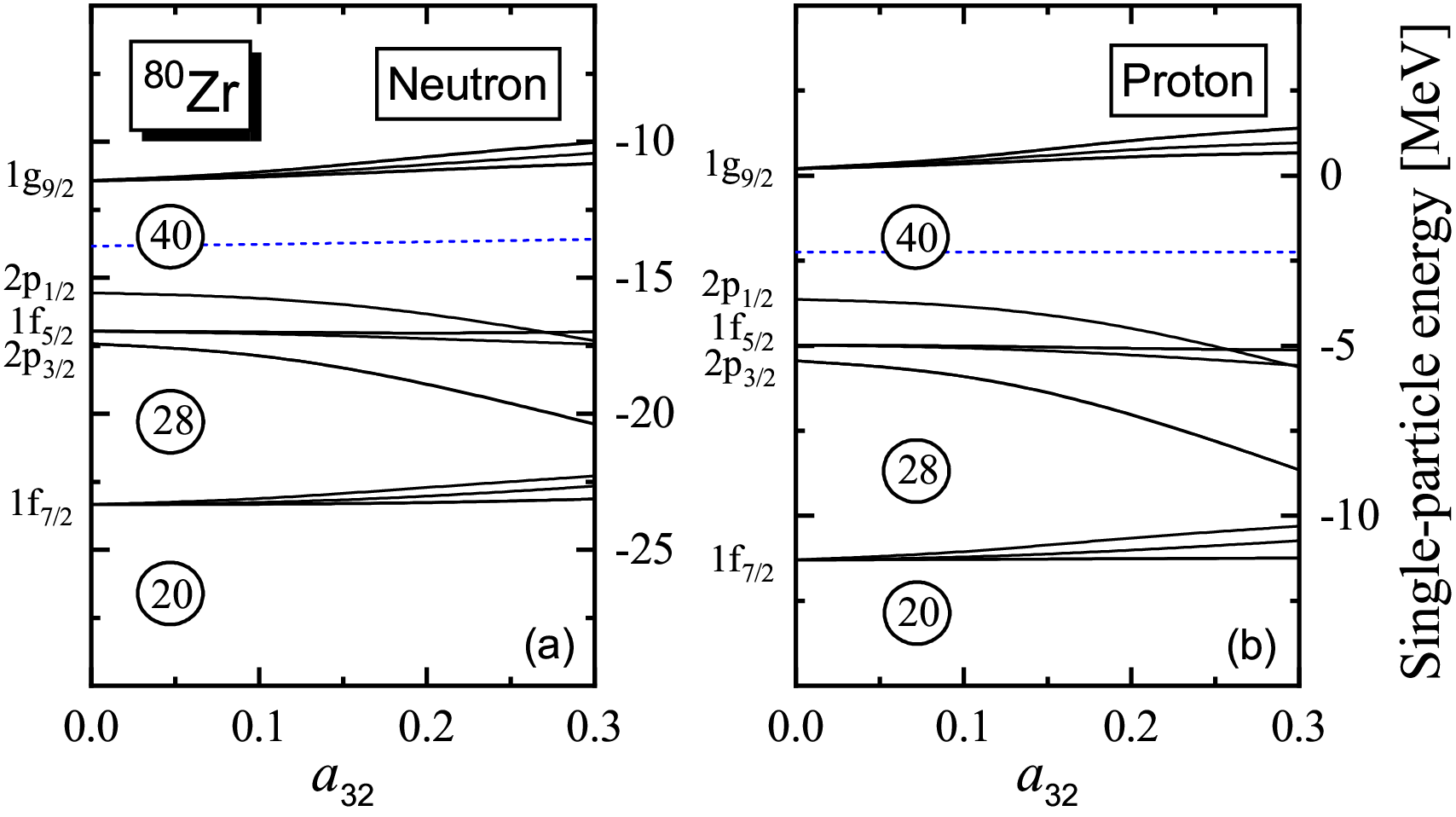}
  \caption{(Color online.) Single-particle energies near the Fermi surface of $^{80}$Zr as a function of the tetrahedral deformation $a_{32}$.
  Levels are labeled by the spherical quantum number of their dominant component, and the dashed lines indicate the Fermi levels.
  }
  \label{Fig3}
\end{figure*}

The tetrahedral deformation in $^{80}$Zr is microscopically explained by the evolution of single-particle levels near the Fermi surface, shown in Fig.~\ref{Fig3}.
At $a_{32}=0$, the single-particle levels are degenerate according to their spherical quantum number $j$.
As $a_{32}$ increases, these levels split into multiplets corresponding to the irreducible representations of the $T_{d}^{D}$ group.
For instance, the spherical $j=5/2$ levels, initially sixfold degenerate, split into a doublet and a quartet as required by tetrahedral symmetry.
This splitting opens pronounced energy gaps at the Fermi surface, lowering the total energy and favoring a finite $a_{32}$, which accounts for the energy gain of $\sim0.84$ MeV observed in the PES (Fig.~\ref{Fig2}).
As a result, the tetrahedral configuration directly drives the anomalous trend of $\delta V_{pn}^{(3)}$ in $^{80}$Zr and $^{82}$Zr (Fig.~\ref{Fig1}), providing a clear microscopic origin for this feature.

\begin{figure*}[!htbp]
  \centering
  \includegraphics[width=0.95\textwidth]{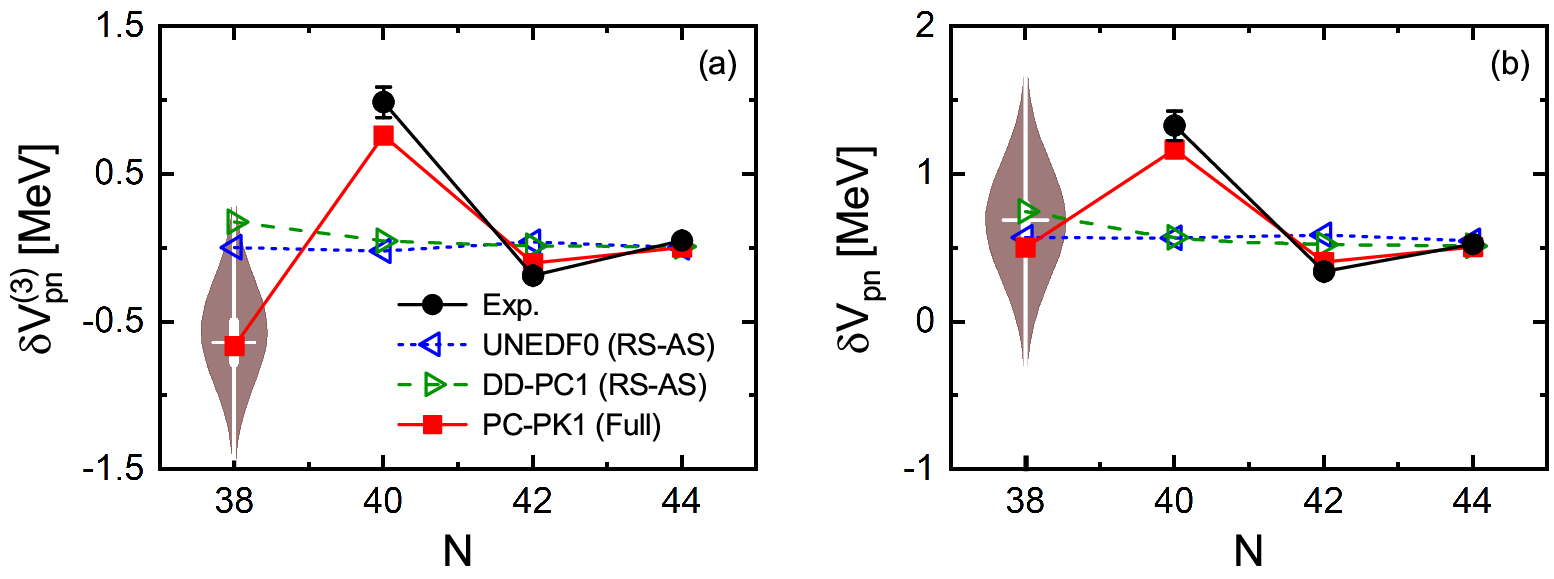}
  \caption{(Color online.) Calculated $\delta V_{pn}^{(3)}$ and $\delta V_{pn}$ for even-even $^{78\text{--}84}$Zr isotopes, compared with experimental data (solid circles) from the AME2020 mass evaluation~\cite{Wang2021CPC}, including extrapolated values.
  The data of $^{80}$Zr is extracted from the recent high precision mass measurement~\cite{Hamaker2021Nature}.
  For $^{78}$Zr, the ground-state binding energy of $^{76}$Zr is not available in AME2020; therefore, predictions from various theoretical models are employed.
  The mean value (horizontal lines) is adopted, and the associated uncertainty is evaluated assuming a Gaussian distribution, which is propagated to the resulting $\delta V_{pn}^{(3)}$ and $\delta V_{pn}$ values and indicated by the shaded area.
  Thick error bars represent uncertainties propagated from the AME2020 data, while thin error bars denote the 99\% credible intervals.
  Results obtained with UNEDF0~\cite{Erler2012Nature} and DD-PC1~\cite{Agbemava2014PRC} under reflection and axial symmetries (RS-AS) are shown by open left and right triangles, respectively, whereas calculations with PC-PK1 allowing full deformation (Full) are represented by solid squares.
  }
  \label{Fig4}
\end{figure*}

As shown above, the tetrahedral deformation in $^{80}$Zr gives rise to the anomalous behavior of $\delta V_{pn}^{(3)}$ in $^{80}$Zr and $^{82}$Zr, establishing $\delta V_{pn}^{(3)}$ as a sensitive indicator of this exotic shape.
This conclusion is further supported by independent density functional calculations and can be tested experimentally through future mass measurements of neutron-deficient Zr isotopes.

Figure~\ref{Fig4}(a) compares $\delta V_{pn}^{(3)}$ values calculated with UNEDF0 and DD-PC1 functionals under reflection and axial symmetries (RS-AS) to the present PC-PK1 results allowing tetrahedral deformation.
Experimental values~\cite{Wang2021CPC}, including that for $^{78}$Zr, are also shown.
The evaluation of $\delta V_{pn}^{(3)}$ for $^{78}$Zr requires the ground-state binding energy of $^{76}$Zr, which has not been measured.
To estimate this quantity, binding energies from a broad set of theoretical models are employed, including WS4+KRR~\cite{Wu2020PRC}, FRDM~\cite{Moller2016ADNDT}, nonrelativistic DFTs (UNEDF0, UNEDF1, SLy4, SV-min)~\cite{Erler2012Nature}, relativistic DFTs (DD-ME2, DD-ME${\rm\delta}$, DD-PC1, NL3$^{*}$)~\cite{Agbemava2014PRC}, and the present 3D lattice PC-PK1 calculation.
The mean of these predictions is adopted, and the associated uncertainty, assumed to follow a Gaussian distribution, is propagated to the resulting $\delta V_{pn}^{(3)}$ value for $^{78}$Zr (shaded area in Fig.~\ref{Fig4}).

As seen in Fig.~\ref{Fig4}(a), neither UNEDF0 nor DD-PC1 reproduces the pronounced peak at $^{80}$Zr due to the absence of tetrahedral deformation, similar to the reflection-symmetric PC-PK1 results [see Fig.~\ref{Fig1}(b)].
This demonstrates the robustness of the $\delta V_{pn}^{(3)}$ indicator across density functionals.
For the most neutron-deficient isotope $^{78}$Zr, the $\delta V_{pn}^{(3)}$ values with reflection symmetry lie at the edge of the estimated uncertainty, whereas calculations allowing full symmetry breaking agree with the central value.
This divergent behavior directly originates from the tetrahedral deformation in $^{80}$Zr.
Consequently, a future experimental determination of the ground-state binding energy of $^{76}$Zr would enable a direct extraction of $\delta V_{pn}^{(3)}$ for $^{78}$Zr, providing a decisive test of the tetrahedral structure in $^{80}$Zr.
Even if the measured value deviates from the present central estimate but remains within the 99\% credible interval, the peak at $^{80}$Zr persists, underscoring the robustness of this indicator.

Figure~\ref{Fig4}(b) shows the corresponding $\delta V_{pn}$ values.
A peak at $^{80}$Zr is observed, though less pronounced, especially considering the 99\% credible interval of $\delta V_{pn}$ for $^{78}$Zr.
While $\delta V_{pn}$ has not previously been recognized as a signature of tetrahedral deformation, it exhibits a local enhancement at $^{80}$Zr.
Since tetrahedral deformation in $^{80}$Zr minimally affects $\delta V_{pn}$ for $^{78}$Zr, the RS-AS and 3D lattice calculations yield nearly identical results.
In contrast, $\delta V_{pn}^{(3)}$ more clearly highlights the effect of tetrahedral deformation, establishing it as a more selective and sensitive probe, while $\delta V_{pn}$ also reflects the underlying structure to a lesser degree.

In summary, we show that the triple binding energy difference $\delta V_{pn}^{(3)}$ constitutes a sensitive and selective probe of tetrahedral deformation in atomic nuclei.
Although the experimental observable alone does not constitute a unique or exclusive proof of tetrahedral symmetry, its characteristic behavior near tetrahedral magic numbers, together with the density functional theory analysis, provides strong evidence for tetrahedral correlations in the relevant Zr nuclei.
The experimental $\delta V_{pn}^{(3)}$ systematics of even-even $^{80\text{--}90}$Zr isotopes are well reproduced without adjustable parameters using relativistic density functional theory on a three-dimensional lattice without any symmetry restrictions.
In particular, the anomalous behavior in $^{80}$Zr and $^{82}$Zr is traced to a well-localized tetrahedral minimum, as supported by potential energy surfaces and single-particle level splittings.
Quadrupole deformation alone cannot account for this anomaly, and calculations with independent density functionals under reflection and axial symmetries similarly fail to reproduce it, highlighting that the effect is robust with respect to the choice of functional but requires tetrahedral deformation to emerge.
Our findings identify $\delta V_{pn}^{(3)}$ as a mass-based and experimentally accessible probe of nuclear tetrahedral symmetry.
Future precision mass measurements of neutron-deficient Zr isotopes, particularly $^{76}$Zr, will provide a decisive test of this prediction and open a new pathway for exploring exotic nonaxial shapes in atomic nuclei.

\begin{acknowledgments}
  We thank Y. M. Xing and X. Xu for helpful discussions.
  This work was partly supported by the National Key Research and Development Program of China 2024YFA1612600, the Fundamental Research Funds for the Central Universities (Grant No. 63261072), and the National Natural Science Foundation of China (Grants No. 12475117, No. 12435006, and No. 12141501), and the High-performance Computing Platform of Peking University.
\end{acknowledgments}

%\bibliography{myreference}

\begin{thebibliography}{64}%
\makeatletter
\providecommand \@ifxundefined [1]{%
 \@ifx{#1\undefined}
}%
\providecommand \@ifnum [1]{%
 \ifnum #1\expandafter \@firstoftwo
 \else \expandafter \@secondoftwo
 \fi
}%
\providecommand \@ifx [1]{%
 \ifx #1\expandafter \@firstoftwo
 \else \expandafter \@secondoftwo
 \fi
}%
\providecommand \natexlab [1]{#1}%
\providecommand \enquote  [1]{``#1''}%
\providecommand \bibnamefont  [1]{#1}%
\providecommand \bibfnamefont [1]{#1}%
\providecommand \citenamefont [1]{#1}%
\providecommand \href@noop [0]{\@secondoftwo}%
\providecommand \href [0]{\begingroup \@sanitize@url \@href}%
\providecommand \@href[1]{\@@startlink{#1}\@@href}%
\providecommand \@@href[1]{\endgroup#1\@@endlink}%
\providecommand \@sanitize@url [0]{\catcode `\\12\catcode `\$12\catcode
  `\&12\catcode `\#12\catcode `\^12\catcode `\_12\catcode `\%12\relax}%
\providecommand \@@startlink[1]{}%
\providecommand \@@endlink[0]{}%
\providecommand \url  [0]{\begingroup\@sanitize@url \@url }%
\providecommand \@url [1]{\endgroup\@href {#1}{\urlprefix }}%
\providecommand \urlprefix  [0]{URL }%
\providecommand \Eprint [0]{\href }%
\providecommand \doibase [0]{http://dx.doi.org/}%
\providecommand \selectlanguage [0]{\@gobble}%
\providecommand \bibinfo  [0]{\@secondoftwo}%
\providecommand \bibfield  [0]{\@secondoftwo}%
\providecommand \translation [1]{[#1]}%
\providecommand \BibitemOpen [0]{}%
\providecommand \bibitemStop [0]{}%
\providecommand \bibitemNoStop [0]{.\EOS\space}%
\providecommand \EOS [0]{\spacefactor3000\relax}%
\providecommand \BibitemShut  [1]{\csname bibitem#1\endcsname}%
\let\auto@bib@innerbib\@empty
%</preamble>
\bibitem [{\citenamefont {Hamamoto}\ \emph {et~al.}(1991)\citenamefont
  {Hamamoto}, \citenamefont {Mottelson}, \citenamefont {Xie},\ and\
  \citenamefont {Zhang}}]{Hamamoto1991ZPD}%
  \BibitemOpen
  \bibfield  {author} {\bibinfo {author} {\bibfnamefont {I.}~\bibnamefont
  {Hamamoto}}, \bibinfo {author} {\bibfnamefont {B.}~\bibnamefont {Mottelson}},
  \bibinfo {author} {\bibfnamefont {H.}~\bibnamefont {Xie}}, \ and\ \bibinfo
  {author} {\bibfnamefont {X.~Z.}\ \bibnamefont {Zhang}},\ }\href {\doibase
  10.1007/BF01425595} {\bibfield  {journal} {\bibinfo  {journal} {Z. Phys. D}\
  }\textbf {\bibinfo {volume} {21}},\ \bibinfo {pages} {163} (\bibinfo {year}
  {1991})}\BibitemShut {NoStop}%
\bibitem [{\citenamefont {Dudek}\ \emph {et~al.}(2002)\citenamefont {Dudek},
  \citenamefont {Go\ifmmode \acute{z}\else \'{z}\fi{}d\ifmmode~\acute{z}\else
  \'{z}\fi{}}, \citenamefont {Schunck},\ and\ \citenamefont
  {Mi\ifmmode~\acute{s}\else \'{s}\fi{}kiewicz}}]{Dudek2002PRL}%
  \BibitemOpen
  \bibfield  {author} {\bibinfo {author} {\bibfnamefont {J.}~\bibnamefont
  {Dudek}}, \bibinfo {author} {\bibfnamefont {A.}~\bibnamefont {Go\ifmmode
  \acute{z}\else \'{z}\fi{}d\ifmmode~\acute{z}\else \'{z}\fi{}}}, \bibinfo
  {author} {\bibfnamefont {N.}~\bibnamefont {Schunck}}, \ and\ \bibinfo
  {author} {\bibfnamefont {M.}~\bibnamefont {Mi\ifmmode~\acute{s}\else
  \'{s}\fi{}kiewicz}},\ }\href {\doibase 10.1103/PhysRevLett.88.252502}
  {\bibfield  {journal} {\bibinfo  {journal} {Phys. Rev. Lett.}\ }\textbf
  {\bibinfo {volume} {88}},\ \bibinfo {pages} {252502} (\bibinfo {year}
  {2002})}\BibitemShut {NoStop}%
\bibitem [{\citenamefont {Bark}\ \emph {et~al.}(2010)\citenamefont {Bark},
  \citenamefont {Sharpey-Schafer}, \citenamefont {Maliage}, \citenamefont
  {Madiba}, \citenamefont {Komati}, \citenamefont {Lawrie}, \citenamefont
  {Lawrie}, \citenamefont {Lindsay}, \citenamefont {Maine}, \citenamefont
  {Mullins}, \citenamefont {Murray}, \citenamefont {Ncapayi}, \citenamefont
  {Ramashidza}, \citenamefont {Smit},\ and\ \citenamefont
  {Vymers}}]{Bark2010PRL}%
  \BibitemOpen
  \bibfield  {author} {\bibinfo {author} {\bibfnamefont {R.~A.}\ \bibnamefont
  {Bark}}, \bibinfo {author} {\bibfnamefont {J.~F.}\ \bibnamefont
  {Sharpey-Schafer}}, \bibinfo {author} {\bibfnamefont {S.~M.}\ \bibnamefont
  {Maliage}}, \bibinfo {author} {\bibfnamefont {T.~E.}\ \bibnamefont {Madiba}},
  \bibinfo {author} {\bibfnamefont {F.~S.}\ \bibnamefont {Komati}}, \bibinfo
  {author} {\bibfnamefont {E.~A.}\ \bibnamefont {Lawrie}}, \bibinfo {author}
  {\bibfnamefont {J.~J.}\ \bibnamefont {Lawrie}}, \bibinfo {author}
  {\bibfnamefont {R.}~\bibnamefont {Lindsay}}, \bibinfo {author} {\bibfnamefont
  {P.}~\bibnamefont {Maine}}, \bibinfo {author} {\bibfnamefont {S.~M.}\
  \bibnamefont {Mullins}}, \bibinfo {author} {\bibfnamefont {S.~H.~T.}\
  \bibnamefont {Murray}}, \bibinfo {author} {\bibfnamefont {N.~J.}\
  \bibnamefont {Ncapayi}}, \bibinfo {author} {\bibfnamefont {T.~M.}\
  \bibnamefont {Ramashidza}}, \bibinfo {author} {\bibfnamefont {F.~D.}\
  \bibnamefont {Smit}}, \ and\ \bibinfo {author} {\bibfnamefont
  {P.}~\bibnamefont {Vymers}},\ }\href {\doibase
  10.1103/PhysRevLett.104.022501} {\bibfield  {journal} {\bibinfo  {journal}
  {Phys. Rev. Lett.}\ }\textbf {\bibinfo {volume} {104}},\ \bibinfo {pages}
  {022501} (\bibinfo {year} {2010})}\BibitemShut {NoStop}%
\bibitem [{\citenamefont {Jentschel}\ \emph {et~al.}(2010)\citenamefont
  {Jentschel}, \citenamefont {Urban}, \citenamefont {Krempel}, \citenamefont
  {Tonev}, \citenamefont {Dudek}, \citenamefont {Curien}, \citenamefont
  {Lauss}, \citenamefont {de~Angelis},\ and\ \citenamefont
  {Petkov}}]{Jentschel2010PRL}%
  \BibitemOpen
  \bibfield  {author} {\bibinfo {author} {\bibfnamefont {M.}~\bibnamefont
  {Jentschel}}, \bibinfo {author} {\bibfnamefont {W.}~\bibnamefont {Urban}},
  \bibinfo {author} {\bibfnamefont {J.}~\bibnamefont {Krempel}}, \bibinfo
  {author} {\bibfnamefont {D.}~\bibnamefont {Tonev}}, \bibinfo {author}
  {\bibfnamefont {J.}~\bibnamefont {Dudek}}, \bibinfo {author} {\bibfnamefont
  {D.}~\bibnamefont {Curien}}, \bibinfo {author} {\bibfnamefont
  {B.}~\bibnamefont {Lauss}}, \bibinfo {author} {\bibfnamefont
  {G.}~\bibnamefont {de~Angelis}}, \ and\ \bibinfo {author} {\bibfnamefont
  {P.}~\bibnamefont {Petkov}},\ }\href {\doibase
  10.1103/PhysRevLett.104.222502} {\bibfield  {journal} {\bibinfo  {journal}
  {Phys. Rev. Lett.}\ }\textbf {\bibinfo {volume} {104}},\ \bibinfo {pages}
  {222502} (\bibinfo {year} {2010})}\BibitemShut {NoStop}%
\bibitem [{\citenamefont {Li}\ and\ \citenamefont {Dudek}(1994)}]{Li1994PRC}%
  \BibitemOpen
  \bibfield  {author} {\bibinfo {author} {\bibfnamefont {X.}~\bibnamefont
  {Li}}\ and\ \bibinfo {author} {\bibfnamefont {J.}~\bibnamefont {Dudek}},\
  }\href {\doibase 10.1103/PhysRevC.49.R1250} {\bibfield  {journal} {\bibinfo
  {journal} {Phys. Rev. C}\ }\textbf {\bibinfo {volume} {49}},\ \bibinfo
  {pages} {R1250} (\bibinfo {year} {1994})}\BibitemShut {NoStop}%
\bibitem [{\citenamefont {Heiss}\ \emph {et~al.}(1999)\citenamefont {Heiss},
  \citenamefont {Lynch},\ and\ \citenamefont {Nazmitdinov}}]{Heiss1999PRC}%
  \BibitemOpen
  \bibfield  {author} {\bibinfo {author} {\bibfnamefont {W.~D.}\ \bibnamefont
  {Heiss}}, \bibinfo {author} {\bibfnamefont {R.~A.}\ \bibnamefont {Lynch}}, \
  and\ \bibinfo {author} {\bibfnamefont {R.~G.}\ \bibnamefont {Nazmitdinov}},\
  }\href {\doibase 10.1103/PhysRevC.60.034303} {\bibfield  {journal} {\bibinfo
  {journal} {Phys. Rev. C}\ }\textbf {\bibinfo {volume} {60}},\ \bibinfo
  {pages} {034303} (\bibinfo {year} {1999})}\BibitemShut {NoStop}%
\bibitem [{\citenamefont {Dudek}\ \emph {et~al.}(2006)\citenamefont {Dudek},
  \citenamefont {Curien}, \citenamefont {Dubray}, \citenamefont {Dobaczewski},
  \citenamefont {Pangon}, \citenamefont {Olbratowski},\ and\ \citenamefont
  {Schunck}}]{Dudek2006PRL}%
  \BibitemOpen
  \bibfield  {author} {\bibinfo {author} {\bibfnamefont {J.}~\bibnamefont
  {Dudek}}, \bibinfo {author} {\bibfnamefont {D.}~\bibnamefont {Curien}},
  \bibinfo {author} {\bibfnamefont {N.}~\bibnamefont {Dubray}}, \bibinfo
  {author} {\bibfnamefont {J.}~\bibnamefont {Dobaczewski}}, \bibinfo {author}
  {\bibfnamefont {V.}~\bibnamefont {Pangon}}, \bibinfo {author} {\bibfnamefont
  {P.}~\bibnamefont {Olbratowski}}, \ and\ \bibinfo {author} {\bibfnamefont
  {N.}~\bibnamefont {Schunck}},\ }\href {\doibase
  10.1103/PhysRevLett.97.072501} {\bibfield  {journal} {\bibinfo  {journal}
  {Phys. Rev. Lett.}\ }\textbf {\bibinfo {volume} {97}},\ \bibinfo {pages}
  {072501} (\bibinfo {year} {2006})}\BibitemShut {NoStop}%
\bibitem [{\citenamefont {Arita}\ and\ \citenamefont
  {Mukumoto}(2014)}]{Arita2014PRC}%
  \BibitemOpen
  \bibfield  {author} {\bibinfo {author} {\bibfnamefont {K.-i.}\ \bibnamefont
  {Arita}}\ and\ \bibinfo {author} {\bibfnamefont {Y.}~\bibnamefont
  {Mukumoto}},\ }\href {\doibase 10.1103/PhysRevC.89.054308} {\bibfield
  {journal} {\bibinfo  {journal} {Phys. Rev. C}\ }\textbf {\bibinfo {volume}
  {89}},\ \bibinfo {pages} {054308} (\bibinfo {year} {2014})}\BibitemShut
  {NoStop}%
\bibitem [{\citenamefont {Dudek}\ \emph {et~al.}(2018)\citenamefont {Dudek},
  \citenamefont {Curien}, \citenamefont {Dedes}, \citenamefont {Mazurek},
  \citenamefont {Tagami}, \citenamefont {Shimizu},\ and\ \citenamefont
  {Bhattacharjee}}]{Dudek2018PRC}%
  \BibitemOpen
  \bibfield  {author} {\bibinfo {author} {\bibfnamefont {J.}~\bibnamefont
  {Dudek}}, \bibinfo {author} {\bibfnamefont {D.}~\bibnamefont {Curien}},
  \bibinfo {author} {\bibfnamefont {I.}~\bibnamefont {Dedes}}, \bibinfo
  {author} {\bibfnamefont {K.}~\bibnamefont {Mazurek}}, \bibinfo {author}
  {\bibfnamefont {S.}~\bibnamefont {Tagami}}, \bibinfo {author} {\bibfnamefont
  {Y.~R.}\ \bibnamefont {Shimizu}}, \ and\ \bibinfo {author} {\bibfnamefont
  {T.}~\bibnamefont {Bhattacharjee}},\ }\href {\doibase
  10.1103/PhysRevC.97.021302} {\bibfield  {journal} {\bibinfo  {journal} {Phys.
  Rev. C}\ }\textbf {\bibinfo {volume} {97}},\ \bibinfo {pages} {021302(R)}
  (\bibinfo {year} {2018})}\BibitemShut {NoStop}%
\bibitem [{\citenamefont {Schunck}\ \emph {et~al.}(2004)\citenamefont
  {Schunck}, \citenamefont {Dudek}, \citenamefont {G\'o\ifmmode \acute{z}\else
  \'{z}\fi{}d\ifmmode~\acute{z}\else \'{z}\fi{}},\ and\ \citenamefont
  {Regan}}]{Schunck2004PRC}%
  \BibitemOpen
  \bibfield  {author} {\bibinfo {author} {\bibfnamefont {N.}~\bibnamefont
  {Schunck}}, \bibinfo {author} {\bibfnamefont {J.}~\bibnamefont {Dudek}},
  \bibinfo {author} {\bibfnamefont {A.}~\bibnamefont {G\'o\ifmmode
  \acute{z}\else \'{z}\fi{}d\ifmmode~\acute{z}\else \'{z}\fi{}}}, \ and\
  \bibinfo {author} {\bibfnamefont {P.~H.}\ \bibnamefont {Regan}},\ }\href
  {\doibase 10.1103/PhysRevC.69.061305} {\bibfield  {journal} {\bibinfo
  {journal} {Phys. Rev. C}\ }\textbf {\bibinfo {volume} {69}},\ \bibinfo
  {pages} {061305(R)} (\bibinfo {year} {2004})}\BibitemShut {NoStop}%
\bibitem [{\citenamefont {Zberecki}\ \emph {et~al.}(2009)\citenamefont
  {Zberecki}, \citenamefont {Heenen},\ and\ \citenamefont
  {Magierski}}]{Zberecki2009PRC}%
  \BibitemOpen
  \bibfield  {author} {\bibinfo {author} {\bibfnamefont {K.}~\bibnamefont
  {Zberecki}}, \bibinfo {author} {\bibfnamefont {P.-H.}\ \bibnamefont
  {Heenen}}, \ and\ \bibinfo {author} {\bibfnamefont {P.}~\bibnamefont
  {Magierski}},\ }\href {\doibase 10.1103/PhysRevC.79.014319} {\bibfield
  {journal} {\bibinfo  {journal} {Phys. Rev. C}\ }\textbf {\bibinfo {volume}
  {79}},\ \bibinfo {pages} {014319} (\bibinfo {year} {2009})}\BibitemShut
  {NoStop}%
\bibitem [{\citenamefont {Tagami}\ \emph {et~al.}(2013)\citenamefont {Tagami},
  \citenamefont {Shimizu},\ and\ \citenamefont {Dudek}}]{Tagami2013PRC}%
  \BibitemOpen
  \bibfield  {author} {\bibinfo {author} {\bibfnamefont {S.}~\bibnamefont
  {Tagami}}, \bibinfo {author} {\bibfnamefont {Y.~R.}\ \bibnamefont {Shimizu}},
  \ and\ \bibinfo {author} {\bibfnamefont {J.}~\bibnamefont {Dudek}},\ }\href
  {\doibase 10.1103/PhysRevC.87.054306} {\bibfield  {journal} {\bibinfo
  {journal} {Phys. Rev. C}\ }\textbf {\bibinfo {volume} {87}},\ \bibinfo
  {pages} {054306} (\bibinfo {year} {2013})}\BibitemShut {NoStop}%
\bibitem [{\citenamefont {Miyahara}\ and\ \citenamefont
  {Nakada}(2018)}]{Miyahara2018PRC}%
  \BibitemOpen
  \bibfield  {author} {\bibinfo {author} {\bibfnamefont {S.}~\bibnamefont
  {Miyahara}}\ and\ \bibinfo {author} {\bibfnamefont {H.}~\bibnamefont
  {Nakada}},\ }\href {\doibase 10.1103/PhysRevC.98.064318} {\bibfield
  {journal} {\bibinfo  {journal} {Phys. Rev. C}\ }\textbf {\bibinfo {volume}
  {98}},\ \bibinfo {pages} {064318} (\bibinfo {year} {2018})}\BibitemShut
  {NoStop}%
\bibitem [{\citenamefont {Wang}\ \emph {et~al.}(2019)\citenamefont {Wang},
  \citenamefont {Dong}, \citenamefont {Gao}, \citenamefont {Chen},\ and\
  \citenamefont {Shen}}]{Wang2019PLB}%
  \BibitemOpen
  \bibfield  {author} {\bibinfo {author} {\bibfnamefont {X.}~\bibnamefont
  {Wang}}, \bibinfo {author} {\bibfnamefont {G.}~\bibnamefont {Dong}}, \bibinfo
  {author} {\bibfnamefont {Z.}~\bibnamefont {Gao}}, \bibinfo {author}
  {\bibfnamefont {Y.}~\bibnamefont {Chen}}, \ and\ \bibinfo {author}
  {\bibfnamefont {C.}~\bibnamefont {Shen}},\ }\href {\doibase
  https://doi.org/10.1016/j.physletb.2019.02.001} {\bibfield  {journal}
  {\bibinfo  {journal} {Phys. Lett. B}\ }\textbf {\bibinfo {volume} {790}},\
  \bibinfo {pages} {498} (\bibinfo {year} {2019})}\BibitemShut {NoStop}%
\bibitem [{\citenamefont {Zhao}\ \emph {et~al.}(2017)\citenamefont {Zhao},
  \citenamefont {Lu}, \citenamefont {Zhao},\ and\ \citenamefont
  {Zhou}}]{Zhao2017PRC}%
  \BibitemOpen
  \bibfield  {author} {\bibinfo {author} {\bibfnamefont {J.}~\bibnamefont
  {Zhao}}, \bibinfo {author} {\bibfnamefont {B.-N.}\ \bibnamefont {Lu}},
  \bibinfo {author} {\bibfnamefont {E.-G.}\ \bibnamefont {Zhao}}, \ and\
  \bibinfo {author} {\bibfnamefont {S.-G.}\ \bibnamefont {Zhou}},\ }\href
  {\doibase 10.1103/PhysRevC.95.014320} {\bibfield  {journal} {\bibinfo
  {journal} {Phys. Rev. C}\ }\textbf {\bibinfo {volume} {95}},\ \bibinfo
  {pages} {014320} (\bibinfo {year} {2017})}\BibitemShut {NoStop}%
\bibitem [{\citenamefont {Rong}\ \emph {et~al.}(2023)\citenamefont {Rong},
  \citenamefont {Wu}, \citenamefont {Lu},\ and\ \citenamefont
  {Yao}}]{Rong2023PLB}%
  \BibitemOpen
  \bibfield  {author} {\bibinfo {author} {\bibfnamefont {Y.-T.}\ \bibnamefont
  {Rong}}, \bibinfo {author} {\bibfnamefont {X.-Y.}\ \bibnamefont {Wu}},
  \bibinfo {author} {\bibfnamefont {B.-N.}\ \bibnamefont {Lu}}, \ and\ \bibinfo
  {author} {\bibfnamefont {J.-M.}\ \bibnamefont {Yao}},\ }\href {\doibase
  https://doi.org/10.1016/j.physletb.2023.137896} {\bibfield  {journal}
  {\bibinfo  {journal} {Phys. Lett. B}\ }\textbf {\bibinfo {volume} {840}},\
  \bibinfo {pages} {137896} (\bibinfo {year} {2023})}\BibitemShut {NoStop}%
\bibitem [{\citenamefont {Xu}\ \emph {et~al.}(2024{\natexlab{a}})\citenamefont
  {Xu}, \citenamefont {Li}, \citenamefont {Ren},\ and\ \citenamefont
  {Zhao}}]{Xu2024PRC}%
  \BibitemOpen
  \bibfield  {author} {\bibinfo {author} {\bibfnamefont {F.~F.}\ \bibnamefont
  {Xu}}, \bibinfo {author} {\bibfnamefont {B.}~\bibnamefont {Li}}, \bibinfo
  {author} {\bibfnamefont {Z.~X.}\ \bibnamefont {Ren}}, \ and\ \bibinfo
  {author} {\bibfnamefont {P.~W.}\ \bibnamefont {Zhao}},\ }\href {\doibase
  10.1103/PhysRevC.109.014311} {\bibfield  {journal} {\bibinfo  {journal}
  {Phys. Rev. C}\ }\textbf {\bibinfo {volume} {109}},\ \bibinfo {pages}
  {014311} (\bibinfo {year} {2024}{\natexlab{a}})}\BibitemShut {NoStop}%
\bibitem [{\citenamefont {Xu}\ \emph {et~al.}(2024{\natexlab{b}})\citenamefont
  {Xu}, \citenamefont {Li}, \citenamefont {Ring},\ and\ \citenamefont
  {Zhao}}]{Xu2024PLB}%
  \BibitemOpen
  \bibfield  {author} {\bibinfo {author} {\bibfnamefont {F.~F.}\ \bibnamefont
  {Xu}}, \bibinfo {author} {\bibfnamefont {B.}~\bibnamefont {Li}}, \bibinfo
  {author} {\bibfnamefont {P.}~\bibnamefont {Ring}}, \ and\ \bibinfo {author}
  {\bibfnamefont {P.~W.}\ \bibnamefont {Zhao}},\ }\href {\doibase
  https://doi.org/10.1016/j.physletb.2024.138893} {\bibfield  {journal}
  {\bibinfo  {journal} {Phys. Lett. B}\ }\textbf {\bibinfo {volume} {856}},\
  \bibinfo {pages} {138893} (\bibinfo {year} {2024}{\natexlab{b}})}\BibitemShut
  {NoStop}%
\bibitem [{\citenamefont {Bijker}\ and\ \citenamefont
  {Iachello}(2014)}]{Bijker2014PRL}%
  \BibitemOpen
  \bibfield  {author} {\bibinfo {author} {\bibfnamefont {R.}~\bibnamefont
  {Bijker}}\ and\ \bibinfo {author} {\bibfnamefont {F.}~\bibnamefont
  {Iachello}},\ }\href {\doibase 10.1103/PhysRevLett.112.152501} {\bibfield
  {journal} {\bibinfo  {journal} {Phys. Rev. Lett.}\ }\textbf {\bibinfo
  {volume} {112}},\ \bibinfo {pages} {152501} (\bibinfo {year}
  {2014})}\BibitemShut {NoStop}%
\bibitem [{\citenamefont {Epelbaum}\ \emph {et~al.}(2014)\citenamefont
  {Epelbaum}, \citenamefont {Krebs}, \citenamefont {L\"ahde}, \citenamefont
  {Lee}, \citenamefont {Mei\ss{}ner},\ and\ \citenamefont
  {Rupak}}]{Epelbaum2014PRL}%
  \BibitemOpen
  \bibfield  {author} {\bibinfo {author} {\bibfnamefont {E.}~\bibnamefont
  {Epelbaum}}, \bibinfo {author} {\bibfnamefont {H.}~\bibnamefont {Krebs}},
  \bibinfo {author} {\bibfnamefont {T.~A.}\ \bibnamefont {L\"ahde}}, \bibinfo
  {author} {\bibfnamefont {D.}~\bibnamefont {Lee}}, \bibinfo {author}
  {\bibfnamefont {U.-G.}\ \bibnamefont {Mei\ss{}ner}}, \ and\ \bibinfo {author}
  {\bibfnamefont {G.}~\bibnamefont {Rupak}},\ }\href {\doibase
  10.1103/PhysRevLett.112.102501} {\bibfield  {journal} {\bibinfo  {journal}
  {Phys. Rev. Lett.}\ }\textbf {\bibinfo {volume} {112}},\ \bibinfo {pages}
  {102501} (\bibinfo {year} {2014})}\BibitemShut {NoStop}%
\bibitem [{\citenamefont {Ackermann}\ \emph {et~al.}(1993)\citenamefont
  {Ackermann}, \citenamefont {Baltzer}, \citenamefont {Ensel}, \citenamefont
  {Freitag}, \citenamefont {Grafen}, \citenamefont {Günther}, \citenamefont
  {Herzog}, \citenamefont {Manns}, \citenamefont {Marten-Tölle}, \citenamefont
  {Müller}, \citenamefont {Prinz}, \citenamefont {Romanski}, \citenamefont
  {Tölle}, \citenamefont {deBoer}, \citenamefont {Gollwitzer},\ and\
  \citenamefont {Maier}}]{Ackermann1993NPA}%
  \BibitemOpen
  \bibfield  {author} {\bibinfo {author} {\bibfnamefont {B.}~\bibnamefont
  {Ackermann}}, \bibinfo {author} {\bibfnamefont {H.}~\bibnamefont {Baltzer}},
  \bibinfo {author} {\bibfnamefont {C.}~\bibnamefont {Ensel}}, \bibinfo
  {author} {\bibfnamefont {K.}~\bibnamefont {Freitag}}, \bibinfo {author}
  {\bibfnamefont {V.}~\bibnamefont {Grafen}}, \bibinfo {author} {\bibfnamefont
  {C.}~\bibnamefont {Günther}}, \bibinfo {author} {\bibfnamefont
  {P.}~\bibnamefont {Herzog}}, \bibinfo {author} {\bibfnamefont
  {J.}~\bibnamefont {Manns}}, \bibinfo {author} {\bibfnamefont
  {M.}~\bibnamefont {Marten-Tölle}}, \bibinfo {author} {\bibfnamefont
  {U.}~\bibnamefont {Müller}}, \bibinfo {author} {\bibfnamefont
  {J.}~\bibnamefont {Prinz}}, \bibinfo {author} {\bibfnamefont
  {I.}~\bibnamefont {Romanski}}, \bibinfo {author} {\bibfnamefont
  {R.}~\bibnamefont {Tölle}}, \bibinfo {author} {\bibfnamefont
  {J.}~\bibnamefont {deBoer}}, \bibinfo {author} {\bibfnamefont
  {N.}~\bibnamefont {Gollwitzer}}, \ and\ \bibinfo {author} {\bibfnamefont
  {H.}~\bibnamefont {Maier}},\ }\href {\doibase
  https://doi.org/10.1016/0375-9474(93)90180-6} {\bibfield  {journal} {\bibinfo
   {journal} {Nucl. Phys. A}\ }\textbf {\bibinfo {volume} {559}},\ \bibinfo
  {pages} {61} (\bibinfo {year} {1993})}\BibitemShut {NoStop}%
\bibitem [{\citenamefont {Ntshangase}\ \emph {et~al.}(2010)\citenamefont
  {Ntshangase}, \citenamefont {Bark}, \citenamefont {Aschman}, \citenamefont
  {Bvumbi}, \citenamefont {Datta}, \citenamefont {Davidson}, \citenamefont
  {Dinoko}, \citenamefont {Elbasher}, \citenamefont {Juh\'asz}, \citenamefont
  {Khaleel}, \citenamefont {Krasznahorkay}, \citenamefont {Lawrie},
  \citenamefont {Lawrie}, \citenamefont {Lieder}, \citenamefont {Majola},
  \citenamefont {Masiteng}, \citenamefont {Mohammed}, \citenamefont {Mullins},
  \citenamefont {Nieminen}, \citenamefont {Nyak\'o}, \citenamefont {Papka},
  \citenamefont {Roux}, \citenamefont {Sharpey-Shafer}, \citenamefont
  {Shirinda}, \citenamefont {Stankiewicz}, \citenamefont {Tim\'ar},\ and\
  \citenamefont {Wilson}}]{Ntshangase2010PRC}%
  \BibitemOpen
  \bibfield  {author} {\bibinfo {author} {\bibfnamefont {S.~S.}\ \bibnamefont
  {Ntshangase}}, \bibinfo {author} {\bibfnamefont {R.~A.}\ \bibnamefont
  {Bark}}, \bibinfo {author} {\bibfnamefont {D.~G.}\ \bibnamefont {Aschman}},
  \bibinfo {author} {\bibfnamefont {S.}~\bibnamefont {Bvumbi}}, \bibinfo
  {author} {\bibfnamefont {P.}~\bibnamefont {Datta}}, \bibinfo {author}
  {\bibfnamefont {P.~M.}\ \bibnamefont {Davidson}}, \bibinfo {author}
  {\bibfnamefont {T.~S.}\ \bibnamefont {Dinoko}}, \bibinfo {author}
  {\bibfnamefont {M.~E.~A.}\ \bibnamefont {Elbasher}}, \bibinfo {author}
  {\bibfnamefont {K.}~\bibnamefont {Juh\'asz}}, \bibinfo {author}
  {\bibfnamefont {E.~M.~A.}\ \bibnamefont {Khaleel}}, \bibinfo {author}
  {\bibfnamefont {A.}~\bibnamefont {Krasznahorkay}}, \bibinfo {author}
  {\bibfnamefont {E.~A.}\ \bibnamefont {Lawrie}}, \bibinfo {author}
  {\bibfnamefont {J.~J.}\ \bibnamefont {Lawrie}}, \bibinfo {author}
  {\bibfnamefont {R.~M.}\ \bibnamefont {Lieder}}, \bibinfo {author}
  {\bibfnamefont {S.~N.~T.}\ \bibnamefont {Majola}}, \bibinfo {author}
  {\bibfnamefont {P.~L.}\ \bibnamefont {Masiteng}}, \bibinfo {author}
  {\bibfnamefont {H.}~\bibnamefont {Mohammed}}, \bibinfo {author}
  {\bibfnamefont {S.~M.}\ \bibnamefont {Mullins}}, \bibinfo {author}
  {\bibfnamefont {P.}~\bibnamefont {Nieminen}}, \bibinfo {author}
  {\bibfnamefont {B.~M.}\ \bibnamefont {Nyak\'o}}, \bibinfo {author}
  {\bibfnamefont {P.}~\bibnamefont {Papka}}, \bibinfo {author} {\bibfnamefont
  {D.~G.}\ \bibnamefont {Roux}}, \bibinfo {author} {\bibfnamefont {J.~F.}\
  \bibnamefont {Sharpey-Shafer}}, \bibinfo {author} {\bibfnamefont
  {O.}~\bibnamefont {Shirinda}}, \bibinfo {author} {\bibfnamefont {M.~A.}\
  \bibnamefont {Stankiewicz}}, \bibinfo {author} {\bibfnamefont
  {J.}~\bibnamefont {Tim\'ar}}, \ and\ \bibinfo {author} {\bibfnamefont
  {A.~N.}\ \bibnamefont {Wilson}},\ }\href {\doibase
  10.1103/PhysRevC.82.041305} {\bibfield  {journal} {\bibinfo  {journal} {Phys.
  Rev. C}\ }\textbf {\bibinfo {volume} {82}},\ \bibinfo {pages} {041305(R)}
  (\bibinfo {year} {2010})}\BibitemShut {NoStop}%
\bibitem [{\citenamefont {Sumikama}\ \emph {et~al.}(2011)\citenamefont
  {Sumikama}, \citenamefont {Yoshinaga}, \citenamefont {Watanabe},
  \citenamefont {Nishimura}, \citenamefont {Miyashita}, \citenamefont
  {Yamaguchi}, \citenamefont {Sugimoto}, \citenamefont {Chiba}, \citenamefont
  {Li}, \citenamefont {Baba}, \citenamefont {Berryman}, \citenamefont {Blasi},
  \citenamefont {Bracco}, \citenamefont {Camera}, \citenamefont {Doornenbal},
  \citenamefont {Go}, \citenamefont {Hashimoto}, \citenamefont {Hayakawa},
  \citenamefont {Hinke}, \citenamefont {Ideguchi}, \citenamefont {Isobe},
  \citenamefont {Ito}, \citenamefont {Jenkins}, \citenamefont {Kawada},
  \citenamefont {Kobayashi}, \citenamefont {Kondo}, \citenamefont {Kr\"ucken},
  \citenamefont {Kubono}, \citenamefont {Lorusso}, \citenamefont {Nakano},
  \citenamefont {Kurata-Nishimura}, \citenamefont {Odahara}, \citenamefont
  {Ong}, \citenamefont {Ota}, \citenamefont {Podoly\'ak}, \citenamefont
  {Sakurai}, \citenamefont {Scheit}, \citenamefont {Steiger}, \citenamefont
  {Steppenbeck}, \citenamefont {Takano}, \citenamefont {Takashima},
  \citenamefont {Tajiri}, \citenamefont {Teranishi}, \citenamefont
  {Wakabayashi}, \citenamefont {Walker}, \citenamefont {Wieland},\ and\
  \citenamefont {Yamaguchi}}]{Sumikama2011PRL}%
  \BibitemOpen
  \bibfield  {author} {\bibinfo {author} {\bibfnamefont {T.}~\bibnamefont
  {Sumikama}}, \bibinfo {author} {\bibfnamefont {K.}~\bibnamefont {Yoshinaga}},
  \bibinfo {author} {\bibfnamefont {H.}~\bibnamefont {Watanabe}}, \bibinfo
  {author} {\bibfnamefont {S.}~\bibnamefont {Nishimura}}, \bibinfo {author}
  {\bibfnamefont {Y.}~\bibnamefont {Miyashita}}, \bibinfo {author}
  {\bibfnamefont {K.}~\bibnamefont {Yamaguchi}}, \bibinfo {author}
  {\bibfnamefont {K.}~\bibnamefont {Sugimoto}}, \bibinfo {author}
  {\bibfnamefont {J.}~\bibnamefont {Chiba}}, \bibinfo {author} {\bibfnamefont
  {Z.}~\bibnamefont {Li}}, \bibinfo {author} {\bibfnamefont {H.}~\bibnamefont
  {Baba}}, \bibinfo {author} {\bibfnamefont {J.~S.}\ \bibnamefont {Berryman}},
  \bibinfo {author} {\bibfnamefont {N.}~\bibnamefont {Blasi}}, \bibinfo
  {author} {\bibfnamefont {A.}~\bibnamefont {Bracco}}, \bibinfo {author}
  {\bibfnamefont {F.}~\bibnamefont {Camera}}, \bibinfo {author} {\bibfnamefont
  {P.}~\bibnamefont {Doornenbal}}, \bibinfo {author} {\bibfnamefont
  {S.}~\bibnamefont {Go}}, \bibinfo {author} {\bibfnamefont {T.}~\bibnamefont
  {Hashimoto}}, \bibinfo {author} {\bibfnamefont {S.}~\bibnamefont {Hayakawa}},
  \bibinfo {author} {\bibfnamefont {C.}~\bibnamefont {Hinke}}, \bibinfo
  {author} {\bibfnamefont {E.}~\bibnamefont {Ideguchi}}, \bibinfo {author}
  {\bibfnamefont {T.}~\bibnamefont {Isobe}}, \bibinfo {author} {\bibfnamefont
  {Y.}~\bibnamefont {Ito}}, \bibinfo {author} {\bibfnamefont {D.~G.}\
  \bibnamefont {Jenkins}}, \bibinfo {author} {\bibfnamefont {Y.}~\bibnamefont
  {Kawada}}, \bibinfo {author} {\bibfnamefont {N.}~\bibnamefont {Kobayashi}},
  \bibinfo {author} {\bibfnamefont {Y.}~\bibnamefont {Kondo}}, \bibinfo
  {author} {\bibfnamefont {R.}~\bibnamefont {Kr\"ucken}}, \bibinfo {author}
  {\bibfnamefont {S.}~\bibnamefont {Kubono}}, \bibinfo {author} {\bibfnamefont
  {G.}~\bibnamefont {Lorusso}}, \bibinfo {author} {\bibfnamefont
  {T.}~\bibnamefont {Nakano}}, \bibinfo {author} {\bibfnamefont
  {M.}~\bibnamefont {Kurata-Nishimura}}, \bibinfo {author} {\bibfnamefont
  {A.}~\bibnamefont {Odahara}}, \bibinfo {author} {\bibfnamefont {H.~J.}\
  \bibnamefont {Ong}}, \bibinfo {author} {\bibfnamefont {S.}~\bibnamefont
  {Ota}}, \bibinfo {author} {\bibfnamefont {Z.}~\bibnamefont {Podoly\'ak}},
  \bibinfo {author} {\bibfnamefont {H.}~\bibnamefont {Sakurai}}, \bibinfo
  {author} {\bibfnamefont {H.}~\bibnamefont {Scheit}}, \bibinfo {author}
  {\bibfnamefont {K.}~\bibnamefont {Steiger}}, \bibinfo {author} {\bibfnamefont
  {D.}~\bibnamefont {Steppenbeck}}, \bibinfo {author} {\bibfnamefont
  {S.}~\bibnamefont {Takano}}, \bibinfo {author} {\bibfnamefont
  {A.}~\bibnamefont {Takashima}}, \bibinfo {author} {\bibfnamefont
  {K.}~\bibnamefont {Tajiri}}, \bibinfo {author} {\bibfnamefont
  {T.}~\bibnamefont {Teranishi}}, \bibinfo {author} {\bibfnamefont
  {Y.}~\bibnamefont {Wakabayashi}}, \bibinfo {author} {\bibfnamefont {P.~M.}\
  \bibnamefont {Walker}}, \bibinfo {author} {\bibfnamefont {O.}~\bibnamefont
  {Wieland}}, \ and\ \bibinfo {author} {\bibfnamefont {H.}~\bibnamefont
  {Yamaguchi}},\ }\href {\doibase 10.1103/PhysRevLett.106.202501} {\bibfield
  {journal} {\bibinfo  {journal} {Phys. Rev. Lett.}\ }\textbf {\bibinfo
  {volume} {106}},\ \bibinfo {pages} {202501} (\bibinfo {year}
  {2011})}\BibitemShut {NoStop}%
\bibitem [{\citenamefont {Hartley}\ \emph {et~al.}(2017)\citenamefont
  {Hartley}, \citenamefont {Riedinger}, \citenamefont {Janssens}, \citenamefont
  {Majola}, \citenamefont {Riley}, \citenamefont {Allmond}, \citenamefont
  {Beausang}, \citenamefont {Carpenter}, \citenamefont {Chiara}, \citenamefont
  {Cooper}, \citenamefont {Curien}, \citenamefont {Gall}, \citenamefont
  {Garrett}, \citenamefont {Kondev}, \citenamefont {Kulp}, \citenamefont
  {Lauritsen}, \citenamefont {McCutchan}, \citenamefont {Miller}, \citenamefont
  {Miller}, \citenamefont {Piot}, \citenamefont {Redon}, \citenamefont
  {Sharpey-Schafer}, \citenamefont {Simpson}, \citenamefont {Stefanescu},
  \citenamefont {Wang}, \citenamefont {Werner}, \citenamefont {Wood},
  \citenamefont {Yu}, \citenamefont {Zhu},\ and\ \citenamefont
  {Dudek}}]{Hartley2017PRC}%
  \BibitemOpen
  \bibfield  {author} {\bibinfo {author} {\bibfnamefont {D.~J.}\ \bibnamefont
  {Hartley}}, \bibinfo {author} {\bibfnamefont {L.~L.}\ \bibnamefont
  {Riedinger}}, \bibinfo {author} {\bibfnamefont {R.~V.~F.}\ \bibnamefont
  {Janssens}}, \bibinfo {author} {\bibfnamefont {S.~N.~T.}\ \bibnamefont
  {Majola}}, \bibinfo {author} {\bibfnamefont {M.~A.}\ \bibnamefont {Riley}},
  \bibinfo {author} {\bibfnamefont {J.~M.}\ \bibnamefont {Allmond}}, \bibinfo
  {author} {\bibfnamefont {C.~W.}\ \bibnamefont {Beausang}}, \bibinfo {author}
  {\bibfnamefont {M.~P.}\ \bibnamefont {Carpenter}}, \bibinfo {author}
  {\bibfnamefont {C.~J.}\ \bibnamefont {Chiara}}, \bibinfo {author}
  {\bibfnamefont {N.}~\bibnamefont {Cooper}}, \bibinfo {author} {\bibfnamefont
  {D.}~\bibnamefont {Curien}}, \bibinfo {author} {\bibfnamefont {B.~J.~P.}\
  \bibnamefont {Gall}}, \bibinfo {author} {\bibfnamefont {P.~E.}\ \bibnamefont
  {Garrett}}, \bibinfo {author} {\bibfnamefont {F.~G.}\ \bibnamefont {Kondev}},
  \bibinfo {author} {\bibfnamefont {W.~D.}\ \bibnamefont {Kulp}}, \bibinfo
  {author} {\bibfnamefont {T.}~\bibnamefont {Lauritsen}}, \bibinfo {author}
  {\bibfnamefont {E.~A.}\ \bibnamefont {McCutchan}}, \bibinfo {author}
  {\bibfnamefont {D.}~\bibnamefont {Miller}}, \bibinfo {author} {\bibfnamefont
  {S.}~\bibnamefont {Miller}}, \bibinfo {author} {\bibfnamefont
  {J.}~\bibnamefont {Piot}}, \bibinfo {author} {\bibfnamefont {N.}~\bibnamefont
  {Redon}}, \bibinfo {author} {\bibfnamefont {J.~F.}\ \bibnamefont
  {Sharpey-Schafer}}, \bibinfo {author} {\bibfnamefont {J.}~\bibnamefont
  {Simpson}}, \bibinfo {author} {\bibfnamefont {I.}~\bibnamefont {Stefanescu}},
  \bibinfo {author} {\bibfnamefont {X.}~\bibnamefont {Wang}}, \bibinfo {author}
  {\bibfnamefont {V.}~\bibnamefont {Werner}}, \bibinfo {author} {\bibfnamefont
  {J.~L.}\ \bibnamefont {Wood}}, \bibinfo {author} {\bibfnamefont {C.-H.}\
  \bibnamefont {Yu}}, \bibinfo {author} {\bibfnamefont {S.}~\bibnamefont
  {Zhu}}, \ and\ \bibinfo {author} {\bibfnamefont {J.}~\bibnamefont {Dudek}},\
  }\href {\doibase 10.1103/PhysRevC.95.014321} {\bibfield  {journal} {\bibinfo
  {journal} {Phys. Rev. C}\ }\textbf {\bibinfo {volume} {95}},\ \bibinfo
  {pages} {014321} (\bibinfo {year} {2017})}\BibitemShut {NoStop}%
\bibitem [{\citenamefont {Zhang}\ \emph {et~al.}(1989)\citenamefont {Zhang},
  \citenamefont {Casten},\ and\ \citenamefont {Brenner}}]{Zhang1989PLB}%
  \BibitemOpen
  \bibfield  {author} {\bibinfo {author} {\bibfnamefont {J.~Y.}\ \bibnamefont
  {Zhang}}, \bibinfo {author} {\bibfnamefont {R.~F.}\ \bibnamefont {Casten}}, \
  and\ \bibinfo {author} {\bibfnamefont {D.~S.}\ \bibnamefont {Brenner}},\
  }\href {\doibase https://doi.org/10.1016/0370-2693(89)91273-2} {\bibfield
  {journal} {\bibinfo  {journal} {Phys. Lett. B}\ }\textbf {\bibinfo {volume}
  {227}},\ \bibinfo {pages} {1} (\bibinfo {year} {1989})}\BibitemShut {NoStop}%
\bibitem [{\citenamefont {Brenner}\ \emph {et~al.}(1990)\citenamefont
  {Brenner}, \citenamefont {Wesselborg}, \citenamefont {Casten}, \citenamefont
  {Warner},\ and\ \citenamefont {Zhang}}]{Brenner1990PLB}%
  \BibitemOpen
  \bibfield  {author} {\bibinfo {author} {\bibfnamefont {D.~S.}\ \bibnamefont
  {Brenner}}, \bibinfo {author} {\bibfnamefont {C.}~\bibnamefont {Wesselborg}},
  \bibinfo {author} {\bibfnamefont {R.~F.}\ \bibnamefont {Casten}}, \bibinfo
  {author} {\bibfnamefont {D.~D.}\ \bibnamefont {Warner}}, \ and\ \bibinfo
  {author} {\bibfnamefont {J.~Y.}\ \bibnamefont {Zhang}},\ }\href {\doibase
  https://doi.org/10.1016/0370-2693(90)90945-3} {\bibfield  {journal} {\bibinfo
   {journal} {Phys. Lett. B}\ }\textbf {\bibinfo {volume} {243}},\ \bibinfo
  {pages} {1} (\bibinfo {year} {1990})}\BibitemShut {NoStop}%
\bibitem [{\citenamefont {Casten}(1985)}]{Casten1985PRL}%
  \BibitemOpen
  \bibfield  {author} {\bibinfo {author} {\bibfnamefont {R.~F.}\ \bibnamefont
  {Casten}},\ }\href {\doibase 10.1103/PhysRevLett.54.1991} {\bibfield
  {journal} {\bibinfo  {journal} {Phys. Rev. Lett.}\ }\textbf {\bibinfo
  {volume} {54}},\ \bibinfo {pages} {1991} (\bibinfo {year}
  {1985})}\BibitemShut {NoStop}%
\bibitem [{\citenamefont {Cakirli}\ and\ \citenamefont
  {Casten}(2006)}]{Cakirli2006PRL}%
  \BibitemOpen
  \bibfield  {author} {\bibinfo {author} {\bibfnamefont {R.~B.}\ \bibnamefont
  {Cakirli}}\ and\ \bibinfo {author} {\bibfnamefont {R.~F.}\ \bibnamefont
  {Casten}},\ }\href {\doibase 10.1103/PhysRevLett.96.132501} {\bibfield
  {journal} {\bibinfo  {journal} {Phys. Rev. Lett.}\ }\textbf {\bibinfo
  {volume} {96}},\ \bibinfo {pages} {132501} (\bibinfo {year}
  {2006})}\BibitemShut {NoStop}%
\bibitem [{\citenamefont {Heyde}\ \emph {et~al.}(1985)\citenamefont {Heyde},
  \citenamefont {{Van Isacker}}, \citenamefont {Casten},\ and\ \citenamefont
  {Wood}}]{Heyde1985PLB}%
  \BibitemOpen
  \bibfield  {author} {\bibinfo {author} {\bibfnamefont {K.}~\bibnamefont
  {Heyde}}, \bibinfo {author} {\bibfnamefont {P.}~\bibnamefont {{Van
  Isacker}}}, \bibinfo {author} {\bibfnamefont {R.}~\bibnamefont {Casten}}, \
  and\ \bibinfo {author} {\bibfnamefont {J.}~\bibnamefont {Wood}},\ }\href
  {\doibase https://doi.org/10.1016/0370-2693(85)91575-8} {\bibfield  {journal}
  {\bibinfo  {journal} {Phys. Lett. B}\ }\textbf {\bibinfo {volume} {155}},\
  \bibinfo {pages} {303} (\bibinfo {year} {1985})}\BibitemShut {NoStop}%
\bibitem [{\citenamefont {Cakirli}\ \emph {et~al.}(2005)\citenamefont
  {Cakirli}, \citenamefont {Brenner}, \citenamefont {Casten},\ and\
  \citenamefont {Millman}}]{Cakirli2005PRL}%
  \BibitemOpen
  \bibfield  {author} {\bibinfo {author} {\bibfnamefont {R.~B.}\ \bibnamefont
  {Cakirli}}, \bibinfo {author} {\bibfnamefont {D.~S.}\ \bibnamefont
  {Brenner}}, \bibinfo {author} {\bibfnamefont {R.~F.}\ \bibnamefont {Casten}},
  \ and\ \bibinfo {author} {\bibfnamefont {E.~A.}\ \bibnamefont {Millman}},\
  }\href {\doibase 10.1103/PhysRevLett.94.092501} {\bibfield  {journal}
  {\bibinfo  {journal} {Phys. Rev. Lett.}\ }\textbf {\bibinfo {volume} {94}},\
  \bibinfo {pages} {092501} (\bibinfo {year} {2005})}\BibitemShut {NoStop}%
\bibitem [{\citenamefont {Van~Isacker}\ \emph {et~al.}(1995)\citenamefont
  {Van~Isacker}, \citenamefont {Warner},\ and\ \citenamefont
  {Brenner}}]{Isacker1995PRL}%
  \BibitemOpen
  \bibfield  {author} {\bibinfo {author} {\bibfnamefont {P.}~\bibnamefont
  {Van~Isacker}}, \bibinfo {author} {\bibfnamefont {D.~D.}\ \bibnamefont
  {Warner}}, \ and\ \bibinfo {author} {\bibfnamefont {D.~S.}\ \bibnamefont
  {Brenner}},\ }\href {\doibase 10.1103/PhysRevLett.74.4607} {\bibfield
  {journal} {\bibinfo  {journal} {Phys. Rev. Lett.}\ }\textbf {\bibinfo
  {volume} {74}},\ \bibinfo {pages} {4607} (\bibinfo {year}
  {1995})}\BibitemShut {NoStop}%
\bibitem [{\citenamefont {Satuła}\ \emph {et~al.}(1997)\citenamefont
  {Satuła}, \citenamefont {Dean}, \citenamefont {Gary}, \citenamefont
  {Mizutori},\ and\ \citenamefont {Nazarewicz}}]{Satula1997PLB}%
  \BibitemOpen
  \bibfield  {author} {\bibinfo {author} {\bibfnamefont {W.}~\bibnamefont
  {Satuła}}, \bibinfo {author} {\bibfnamefont {D.}~\bibnamefont {Dean}},
  \bibinfo {author} {\bibfnamefont {J.}~\bibnamefont {Gary}}, \bibinfo {author}
  {\bibfnamefont {S.}~\bibnamefont {Mizutori}}, \ and\ \bibinfo {author}
  {\bibfnamefont {W.}~\bibnamefont {Nazarewicz}},\ }\href {\doibase
  https://doi.org/10.1016/S0370-2693(97)00711-9} {\bibfield  {journal}
  {\bibinfo  {journal} {Phys. Lett. B}\ }\textbf {\bibinfo {volume} {407}},\
  \bibinfo {pages} {103} (\bibinfo {year} {1997})}\BibitemShut {NoStop}%
\bibitem [{\citenamefont {Chen}\ \emph {et~al.}(2009)\citenamefont {Chen},
  \citenamefont {Litvinov}, \citenamefont {Pla\ss{}}, \citenamefont {Beckert},
  \citenamefont {Beller}, \citenamefont {Bosch}, \citenamefont {Boutin},
  \citenamefont {Caceres}, \citenamefont {Cakirli}, \citenamefont {Carroll},
  \citenamefont {Casten}, \citenamefont {Chakrawarthy}, \citenamefont {Cullen},
  \citenamefont {Cullen}, \citenamefont {Franzke}, \citenamefont {Geissel},
  \citenamefont {Gerl}, \citenamefont {G\'orska}, \citenamefont {Jones},
  \citenamefont {Kishada}, \citenamefont {Kn\"obel}, \citenamefont
  {Kozhuharov}, \citenamefont {Litvinov}, \citenamefont {Liu}, \citenamefont
  {Mandal}, \citenamefont {Montes}, \citenamefont {M\"unzenberg}, \citenamefont
  {Nolden}, \citenamefont {Ohtsubo}, \citenamefont {Patyk}, \citenamefont
  {Podoly\'ak}, \citenamefont {Propri}, \citenamefont {Rigby}, \citenamefont
  {Saito}, \citenamefont {Saito}, \citenamefont {Scheidenberger}, \citenamefont
  {Shindo}, \citenamefont {Steck}, \citenamefont {Ugorowski}, \citenamefont
  {Walker}, \citenamefont {Williams}, \citenamefont {Weick}, \citenamefont
  {Winkler}, \citenamefont {Wollersheim},\ and\ \citenamefont
  {Yamaguchi}}]{Chen2009PRL}%
  \BibitemOpen
  \bibfield  {author} {\bibinfo {author} {\bibfnamefont {L.}~\bibnamefont
  {Chen}}, \bibinfo {author} {\bibfnamefont {Y.~A.}\ \bibnamefont {Litvinov}},
  \bibinfo {author} {\bibfnamefont {W.~R.}\ \bibnamefont {Pla\ss{}}}, \bibinfo
  {author} {\bibfnamefont {K.}~\bibnamefont {Beckert}}, \bibinfo {author}
  {\bibfnamefont {P.}~\bibnamefont {Beller}}, \bibinfo {author} {\bibfnamefont
  {F.}~\bibnamefont {Bosch}}, \bibinfo {author} {\bibfnamefont
  {D.}~\bibnamefont {Boutin}}, \bibinfo {author} {\bibfnamefont
  {L.}~\bibnamefont {Caceres}}, \bibinfo {author} {\bibfnamefont {R.~B.}\
  \bibnamefont {Cakirli}}, \bibinfo {author} {\bibfnamefont {J.~J.}\
  \bibnamefont {Carroll}}, \bibinfo {author} {\bibfnamefont {R.~F.}\
  \bibnamefont {Casten}}, \bibinfo {author} {\bibfnamefont {R.~S.}\
  \bibnamefont {Chakrawarthy}}, \bibinfo {author} {\bibfnamefont {D.~M.}\
  \bibnamefont {Cullen}}, \bibinfo {author} {\bibfnamefont {I.~J.}\
  \bibnamefont {Cullen}}, \bibinfo {author} {\bibfnamefont {B.}~\bibnamefont
  {Franzke}}, \bibinfo {author} {\bibfnamefont {H.}~\bibnamefont {Geissel}},
  \bibinfo {author} {\bibfnamefont {J.}~\bibnamefont {Gerl}}, \bibinfo {author}
  {\bibfnamefont {M.}~\bibnamefont {G\'orska}}, \bibinfo {author}
  {\bibfnamefont {G.~A.}\ \bibnamefont {Jones}}, \bibinfo {author}
  {\bibfnamefont {A.}~\bibnamefont {Kishada}}, \bibinfo {author} {\bibfnamefont
  {R.}~\bibnamefont {Kn\"obel}}, \bibinfo {author} {\bibfnamefont
  {C.}~\bibnamefont {Kozhuharov}}, \bibinfo {author} {\bibfnamefont {S.~A.}\
  \bibnamefont {Litvinov}}, \bibinfo {author} {\bibfnamefont {Z.}~\bibnamefont
  {Liu}}, \bibinfo {author} {\bibfnamefont {S.}~\bibnamefont {Mandal}},
  \bibinfo {author} {\bibfnamefont {F.}~\bibnamefont {Montes}}, \bibinfo
  {author} {\bibfnamefont {G.}~\bibnamefont {M\"unzenberg}}, \bibinfo {author}
  {\bibfnamefont {F.}~\bibnamefont {Nolden}}, \bibinfo {author} {\bibfnamefont
  {T.}~\bibnamefont {Ohtsubo}}, \bibinfo {author} {\bibfnamefont
  {Z.}~\bibnamefont {Patyk}}, \bibinfo {author} {\bibfnamefont
  {Z.}~\bibnamefont {Podoly\'ak}}, \bibinfo {author} {\bibfnamefont
  {R.}~\bibnamefont {Propri}}, \bibinfo {author} {\bibfnamefont
  {S.}~\bibnamefont {Rigby}}, \bibinfo {author} {\bibfnamefont
  {N.}~\bibnamefont {Saito}}, \bibinfo {author} {\bibfnamefont
  {T.}~\bibnamefont {Saito}}, \bibinfo {author} {\bibfnamefont
  {C.}~\bibnamefont {Scheidenberger}}, \bibinfo {author} {\bibfnamefont
  {M.}~\bibnamefont {Shindo}}, \bibinfo {author} {\bibfnamefont
  {M.}~\bibnamefont {Steck}}, \bibinfo {author} {\bibfnamefont
  {P.}~\bibnamefont {Ugorowski}}, \bibinfo {author} {\bibfnamefont {P.~M.}\
  \bibnamefont {Walker}}, \bibinfo {author} {\bibfnamefont {S.}~\bibnamefont
  {Williams}}, \bibinfo {author} {\bibfnamefont {H.}~\bibnamefont {Weick}},
  \bibinfo {author} {\bibfnamefont {M.}~\bibnamefont {Winkler}}, \bibinfo
  {author} {\bibfnamefont {H.-J.}\ \bibnamefont {Wollersheim}}, \ and\ \bibinfo
  {author} {\bibfnamefont {T.}~\bibnamefont {Yamaguchi}},\ }\href {\doibase
  10.1103/PhysRevLett.102.122503} {\bibfield  {journal} {\bibinfo  {journal}
  {Phys. Rev. Lett.}\ }\textbf {\bibinfo {volume} {102}},\ \bibinfo {pages}
  {122503} (\bibinfo {year} {2009})}\BibitemShut {NoStop}%
\bibitem [{\citenamefont {Neidherr}\ \emph {et~al.}(2009)\citenamefont
  {Neidherr}, \citenamefont {Audi}, \citenamefont {Beck}, \citenamefont
  {Blaum}, \citenamefont {B\"ohm}, \citenamefont {Breitenfeldt}, \citenamefont
  {Cakirli}, \citenamefont {Casten}, \citenamefont {George}, \citenamefont
  {Herfurth}, \citenamefont {Herlert}, \citenamefont {Kellerbauer},
  \citenamefont {Kowalska}, \citenamefont {Lunney}, \citenamefont
  {Minaya-Ramirez}, \citenamefont {Naimi}, \citenamefont {Noah}, \citenamefont
  {Penescu}, \citenamefont {Rosenbusch}, \citenamefont {Schwarz}, \citenamefont
  {Schweikhard},\ and\ \citenamefont {Stora}}]{Neidherr2009PRL}%
  \BibitemOpen
  \bibfield  {author} {\bibinfo {author} {\bibfnamefont {D.}~\bibnamefont
  {Neidherr}}, \bibinfo {author} {\bibfnamefont {G.}~\bibnamefont {Audi}},
  \bibinfo {author} {\bibfnamefont {D.}~\bibnamefont {Beck}}, \bibinfo {author}
  {\bibfnamefont {K.}~\bibnamefont {Blaum}}, \bibinfo {author} {\bibfnamefont
  {C.}~\bibnamefont {B\"ohm}}, \bibinfo {author} {\bibfnamefont
  {M.}~\bibnamefont {Breitenfeldt}}, \bibinfo {author} {\bibfnamefont {R.~B.}\
  \bibnamefont {Cakirli}}, \bibinfo {author} {\bibfnamefont {R.~F.}\
  \bibnamefont {Casten}}, \bibinfo {author} {\bibfnamefont {S.}~\bibnamefont
  {George}}, \bibinfo {author} {\bibfnamefont {F.}~\bibnamefont {Herfurth}},
  \bibinfo {author} {\bibfnamefont {A.}~\bibnamefont {Herlert}}, \bibinfo
  {author} {\bibfnamefont {A.}~\bibnamefont {Kellerbauer}}, \bibinfo {author}
  {\bibfnamefont {M.}~\bibnamefont {Kowalska}}, \bibinfo {author}
  {\bibfnamefont {D.}~\bibnamefont {Lunney}}, \bibinfo {author} {\bibfnamefont
  {E.}~\bibnamefont {Minaya-Ramirez}}, \bibinfo {author} {\bibfnamefont
  {S.}~\bibnamefont {Naimi}}, \bibinfo {author} {\bibfnamefont
  {E.}~\bibnamefont {Noah}}, \bibinfo {author} {\bibfnamefont {L.}~\bibnamefont
  {Penescu}}, \bibinfo {author} {\bibfnamefont {M.}~\bibnamefont {Rosenbusch}},
  \bibinfo {author} {\bibfnamefont {S.}~\bibnamefont {Schwarz}}, \bibinfo
  {author} {\bibfnamefont {L.}~\bibnamefont {Schweikhard}}, \ and\ \bibinfo
  {author} {\bibfnamefont {T.}~\bibnamefont {Stora}},\ }\href {\doibase
  10.1103/PhysRevLett.102.112501} {\bibfield  {journal} {\bibinfo  {journal}
  {Phys. Rev. Lett.}\ }\textbf {\bibinfo {volume} {102}},\ \bibinfo {pages}
  {112501} (\bibinfo {year} {2009})}\BibitemShut {NoStop}%
\bibitem [{\citenamefont {Ketelaer}\ \emph {et~al.}(2011)\citenamefont
  {Ketelaer}, \citenamefont {Audi}, \citenamefont {Beyer}, \citenamefont
  {Blaum}, \citenamefont {Block}, \citenamefont {Cakirli}, \citenamefont
  {Casten}, \citenamefont {Droese}, \citenamefont {Dworschak}, \citenamefont
  {Eberhardt}, \citenamefont {Eibach}, \citenamefont {Herfurth}, \citenamefont
  {Minaya~Ramirez}, \citenamefont {Nagy}, \citenamefont {Neidherr},
  \citenamefont {N\"ortersh\"auser}, \citenamefont {Smorra},\ and\
  \citenamefont {Wang}}]{Ketelaer2011PRC}%
  \BibitemOpen
  \bibfield  {author} {\bibinfo {author} {\bibfnamefont {J.}~\bibnamefont
  {Ketelaer}}, \bibinfo {author} {\bibfnamefont {G.}~\bibnamefont {Audi}},
  \bibinfo {author} {\bibfnamefont {T.}~\bibnamefont {Beyer}}, \bibinfo
  {author} {\bibfnamefont {K.}~\bibnamefont {Blaum}}, \bibinfo {author}
  {\bibfnamefont {M.}~\bibnamefont {Block}}, \bibinfo {author} {\bibfnamefont
  {R.~B.}\ \bibnamefont {Cakirli}}, \bibinfo {author} {\bibfnamefont {R.~F.}\
  \bibnamefont {Casten}}, \bibinfo {author} {\bibfnamefont {C.}~\bibnamefont
  {Droese}}, \bibinfo {author} {\bibfnamefont {M.}~\bibnamefont {Dworschak}},
  \bibinfo {author} {\bibfnamefont {K.}~\bibnamefont {Eberhardt}}, \bibinfo
  {author} {\bibfnamefont {M.}~\bibnamefont {Eibach}}, \bibinfo {author}
  {\bibfnamefont {F.}~\bibnamefont {Herfurth}}, \bibinfo {author}
  {\bibfnamefont {E.}~\bibnamefont {Minaya~Ramirez}}, \bibinfo {author}
  {\bibfnamefont {S.}~\bibnamefont {Nagy}}, \bibinfo {author} {\bibfnamefont
  {D.}~\bibnamefont {Neidherr}}, \bibinfo {author} {\bibfnamefont
  {W.}~\bibnamefont {N\"ortersh\"auser}}, \bibinfo {author} {\bibfnamefont
  {C.}~\bibnamefont {Smorra}}, \ and\ \bibinfo {author} {\bibfnamefont
  {M.}~\bibnamefont {Wang}},\ }\href {\doibase 10.1103/PhysRevC.84.014311}
  {\bibfield  {journal} {\bibinfo  {journal} {Phys. Rev. C}\ }\textbf {\bibinfo
  {volume} {84}},\ \bibinfo {pages} {014311} (\bibinfo {year}
  {2011})}\BibitemShut {NoStop}%
\bibitem [{\citenamefont {Mardor}\ \emph {et~al.}(2021)\citenamefont {Mardor},
  \citenamefont {Andr\'es}, \citenamefont {Dickel}, \citenamefont {Amanbayev},
  \citenamefont {Beck}, \citenamefont {Bergmann}, \citenamefont {Geissel},
  \citenamefont {Gr\"of}, \citenamefont {Haettner}, \citenamefont {Hornung},
  \citenamefont {Kalantar-Nayestanaki}, \citenamefont {Kripko-Koncz},
  \citenamefont {Miskun}, \citenamefont {Mollaebrahimi}, \citenamefont
  {Pla\ss{}}, \citenamefont {Scheidenberger}, \citenamefont {Weick},
  \citenamefont {Bagchi}, \citenamefont {Balabanski}, \citenamefont {Bezbakh},
  \citenamefont {Brencic}, \citenamefont {Charviakova}, \citenamefont
  {Chudoba}, \citenamefont {Constantin}, \citenamefont {Dehghan}, \citenamefont
  {Fomichev}, \citenamefont {Grigorenko}, \citenamefont {Hall}, \citenamefont
  {Harakeh}, \citenamefont {Hucka}, \citenamefont {Kankainen}, \citenamefont
  {Kiselev}, \citenamefont {Kn\"obel}, \citenamefont {Kostyleva}, \citenamefont
  {Krupko}, \citenamefont {Kurkova}, \citenamefont {Kuzminchuk}, \citenamefont
  {Mukha}, \citenamefont {Muzalevskii}, \citenamefont {Nichita}, \citenamefont
  {Nociforo}, \citenamefont {Patyk}, \citenamefont {Pf\"utzner}, \citenamefont
  {Pietri}, \citenamefont {Purushothaman}, \citenamefont {Reiter},
  \citenamefont {Roesch}, \citenamefont {Schirru}, \citenamefont {Sharov},
  \citenamefont {Sp\ifmmode~\u{a}\else \u{a}\fi{}taru}, \citenamefont {Stanic},
  \citenamefont {State}, \citenamefont {Tanaka}, \citenamefont {Vencelj},
  \citenamefont {Yavor},\ and\ \citenamefont {Zhao}}]{Mardor2021PRC}%
  \BibitemOpen
  \bibfield  {author} {\bibinfo {author} {\bibfnamefont {I.}~\bibnamefont
  {Mardor}}, \bibinfo {author} {\bibfnamefont {S.~A.~S.}\ \bibnamefont
  {Andr\'es}}, \bibinfo {author} {\bibfnamefont {T.}~\bibnamefont {Dickel}},
  \bibinfo {author} {\bibfnamefont {D.}~\bibnamefont {Amanbayev}}, \bibinfo
  {author} {\bibfnamefont {S.}~\bibnamefont {Beck}}, \bibinfo {author}
  {\bibfnamefont {J.}~\bibnamefont {Bergmann}}, \bibinfo {author}
  {\bibfnamefont {H.}~\bibnamefont {Geissel}}, \bibinfo {author} {\bibfnamefont
  {L.}~\bibnamefont {Gr\"of}}, \bibinfo {author} {\bibfnamefont
  {E.}~\bibnamefont {Haettner}}, \bibinfo {author} {\bibfnamefont
  {C.}~\bibnamefont {Hornung}}, \bibinfo {author} {\bibfnamefont
  {N.}~\bibnamefont {Kalantar-Nayestanaki}}, \bibinfo {author} {\bibfnamefont
  {G.}~\bibnamefont {Kripko-Koncz}}, \bibinfo {author} {\bibfnamefont
  {I.}~\bibnamefont {Miskun}}, \bibinfo {author} {\bibfnamefont
  {A.}~\bibnamefont {Mollaebrahimi}}, \bibinfo {author} {\bibfnamefont {W.~R.}\
  \bibnamefont {Pla\ss{}}}, \bibinfo {author} {\bibfnamefont {C.}~\bibnamefont
  {Scheidenberger}}, \bibinfo {author} {\bibfnamefont {H.}~\bibnamefont
  {Weick}}, \bibinfo {author} {\bibfnamefont {S.}~\bibnamefont {Bagchi}},
  \bibinfo {author} {\bibfnamefont {D.~L.}\ \bibnamefont {Balabanski}},
  \bibinfo {author} {\bibfnamefont {A.~A.}\ \bibnamefont {Bezbakh}}, \bibinfo
  {author} {\bibfnamefont {Z.}~\bibnamefont {Brencic}}, \bibinfo {author}
  {\bibfnamefont {O.}~\bibnamefont {Charviakova}}, \bibinfo {author}
  {\bibfnamefont {V.}~\bibnamefont {Chudoba}}, \bibinfo {author} {\bibfnamefont
  {P.}~\bibnamefont {Constantin}}, \bibinfo {author} {\bibfnamefont
  {M.}~\bibnamefont {Dehghan}}, \bibinfo {author} {\bibfnamefont {A.~S.}\
  \bibnamefont {Fomichev}}, \bibinfo {author} {\bibfnamefont {L.~V.}\
  \bibnamefont {Grigorenko}}, \bibinfo {author} {\bibfnamefont
  {O.}~\bibnamefont {Hall}}, \bibinfo {author} {\bibfnamefont {M.~N.}\
  \bibnamefont {Harakeh}}, \bibinfo {author} {\bibfnamefont {J.-P.}\
  \bibnamefont {Hucka}}, \bibinfo {author} {\bibfnamefont {A.}~\bibnamefont
  {Kankainen}}, \bibinfo {author} {\bibfnamefont {O.}~\bibnamefont {Kiselev}},
  \bibinfo {author} {\bibfnamefont {R.}~\bibnamefont {Kn\"obel}}, \bibinfo
  {author} {\bibfnamefont {D.~A.}\ \bibnamefont {Kostyleva}}, \bibinfo {author}
  {\bibfnamefont {S.~A.}\ \bibnamefont {Krupko}}, \bibinfo {author}
  {\bibfnamefont {N.}~\bibnamefont {Kurkova}}, \bibinfo {author} {\bibfnamefont
  {N.}~\bibnamefont {Kuzminchuk}}, \bibinfo {author} {\bibfnamefont
  {I.}~\bibnamefont {Mukha}}, \bibinfo {author} {\bibfnamefont {I.~A.}\
  \bibnamefont {Muzalevskii}}, \bibinfo {author} {\bibfnamefont
  {D.}~\bibnamefont {Nichita}}, \bibinfo {author} {\bibfnamefont
  {C.}~\bibnamefont {Nociforo}}, \bibinfo {author} {\bibfnamefont
  {Z.}~\bibnamefont {Patyk}}, \bibinfo {author} {\bibfnamefont
  {M.}~\bibnamefont {Pf\"utzner}}, \bibinfo {author} {\bibfnamefont
  {S.}~\bibnamefont {Pietri}}, \bibinfo {author} {\bibfnamefont
  {S.}~\bibnamefont {Purushothaman}}, \bibinfo {author} {\bibfnamefont {M.~P.}\
  \bibnamefont {Reiter}}, \bibinfo {author} {\bibfnamefont {H.}~\bibnamefont
  {Roesch}}, \bibinfo {author} {\bibfnamefont {F.}~\bibnamefont {Schirru}},
  \bibinfo {author} {\bibfnamefont {P.~G.}\ \bibnamefont {Sharov}}, \bibinfo
  {author} {\bibfnamefont {A.}~\bibnamefont {Sp\ifmmode~\u{a}\else
  \u{a}\fi{}taru}}, \bibinfo {author} {\bibfnamefont {G.}~\bibnamefont
  {Stanic}}, \bibinfo {author} {\bibfnamefont {A.}~\bibnamefont {State}},
  \bibinfo {author} {\bibfnamefont {Y.~K.}\ \bibnamefont {Tanaka}}, \bibinfo
  {author} {\bibfnamefont {M.}~\bibnamefont {Vencelj}}, \bibinfo {author}
  {\bibfnamefont {M.~I.}\ \bibnamefont {Yavor}}, \ and\ \bibinfo {author}
  {\bibfnamefont {J.}~\bibnamefont {Zhao}},\ }\href {\doibase
  10.1103/PhysRevC.103.034319} {\bibfield  {journal} {\bibinfo  {journal}
  {Phys. Rev. C}\ }\textbf {\bibinfo {volume} {103}},\ \bibinfo {pages}
  {034319} (\bibinfo {year} {2021})}\BibitemShut {NoStop}%
\bibitem [{\citenamefont {Wang}\ \emph {et~al.}(2023)\citenamefont {Wang},
  \citenamefont {Zhang}, \citenamefont {Zhou}, \citenamefont {Zhou},
  \citenamefont {Xu}, \citenamefont {Liu}, \citenamefont {Li}, \citenamefont
  {Niu}, \citenamefont {Huang}, \citenamefont {Yuan}, \citenamefont {Zhang},
  \citenamefont {Xu}, \citenamefont {Litvinov}, \citenamefont {Blaum},
  \citenamefont {Meisel}, \citenamefont {Casten}, \citenamefont {Cakirli},
  \citenamefont {Chen}, \citenamefont {Deng}, \citenamefont {Fu}, \citenamefont
  {Ge}, \citenamefont {Li}, \citenamefont {Liao}, \citenamefont {Litvinov},
  \citenamefont {Shuai}, \citenamefont {Shi}, \citenamefont {Song},
  \citenamefont {Sun}, \citenamefont {Wang}, \citenamefont {Xing},
  \citenamefont {Xu}, \citenamefont {Yan}, \citenamefont {Yang}, \citenamefont
  {Yuan}, \citenamefont {Zeng},\ and\ \citenamefont {Zhang}}]{Wang2023PRL}%
  \BibitemOpen
  \bibfield  {author} {\bibinfo {author} {\bibfnamefont {M.}~\bibnamefont
  {Wang}}, \bibinfo {author} {\bibfnamefont {Y.~H.}\ \bibnamefont {Zhang}},
  \bibinfo {author} {\bibfnamefont {X.}~\bibnamefont {Zhou}}, \bibinfo {author}
  {\bibfnamefont {X.~H.}\ \bibnamefont {Zhou}}, \bibinfo {author}
  {\bibfnamefont {H.~S.}\ \bibnamefont {Xu}}, \bibinfo {author} {\bibfnamefont
  {M.~L.}\ \bibnamefont {Liu}}, \bibinfo {author} {\bibfnamefont {J.~G.}\
  \bibnamefont {Li}}, \bibinfo {author} {\bibfnamefont {Y.~F.}\ \bibnamefont
  {Niu}}, \bibinfo {author} {\bibfnamefont {W.~J.}\ \bibnamefont {Huang}},
  \bibinfo {author} {\bibfnamefont {Q.}~\bibnamefont {Yuan}}, \bibinfo {author}
  {\bibfnamefont {S.}~\bibnamefont {Zhang}}, \bibinfo {author} {\bibfnamefont
  {F.~R.}\ \bibnamefont {Xu}}, \bibinfo {author} {\bibfnamefont {Y.~A.}\
  \bibnamefont {Litvinov}}, \bibinfo {author} {\bibfnamefont {K.}~\bibnamefont
  {Blaum}}, \bibinfo {author} {\bibfnamefont {Z.}~\bibnamefont {Meisel}},
  \bibinfo {author} {\bibfnamefont {R.~F.}\ \bibnamefont {Casten}}, \bibinfo
  {author} {\bibfnamefont {R.~B.}\ \bibnamefont {Cakirli}}, \bibinfo {author}
  {\bibfnamefont {R.~J.}\ \bibnamefont {Chen}}, \bibinfo {author}
  {\bibfnamefont {H.~Y.}\ \bibnamefont {Deng}}, \bibinfo {author}
  {\bibfnamefont {C.~Y.}\ \bibnamefont {Fu}}, \bibinfo {author} {\bibfnamefont
  {W.~W.}\ \bibnamefont {Ge}}, \bibinfo {author} {\bibfnamefont {H.~F.}\
  \bibnamefont {Li}}, \bibinfo {author} {\bibfnamefont {T.}~\bibnamefont
  {Liao}}, \bibinfo {author} {\bibfnamefont {S.~A.}\ \bibnamefont {Litvinov}},
  \bibinfo {author} {\bibfnamefont {P.}~\bibnamefont {Shuai}}, \bibinfo
  {author} {\bibfnamefont {J.~Y.}\ \bibnamefont {Shi}}, \bibinfo {author}
  {\bibfnamefont {Y.~N.}\ \bibnamefont {Song}}, \bibinfo {author}
  {\bibfnamefont {M.~Z.}\ \bibnamefont {Sun}}, \bibinfo {author} {\bibfnamefont
  {Q.}~\bibnamefont {Wang}}, \bibinfo {author} {\bibfnamefont {Y.~M.}\
  \bibnamefont {Xing}}, \bibinfo {author} {\bibfnamefont {X.}~\bibnamefont
  {Xu}}, \bibinfo {author} {\bibfnamefont {X.~L.}\ \bibnamefont {Yan}},
  \bibinfo {author} {\bibfnamefont {J.~C.}\ \bibnamefont {Yang}}, \bibinfo
  {author} {\bibfnamefont {Y.~J.}\ \bibnamefont {Yuan}}, \bibinfo {author}
  {\bibfnamefont {Q.}~\bibnamefont {Zeng}}, \ and\ \bibinfo {author}
  {\bibfnamefont {M.}~\bibnamefont {Zhang}},\ }\href {\doibase
  10.1103/PhysRevLett.130.192501} {\bibfield  {journal} {\bibinfo  {journal}
  {Phys. Rev. Lett.}\ }\textbf {\bibinfo {volume} {130}},\ \bibinfo {pages}
  {192501} (\bibinfo {year} {2023})}\BibitemShut {NoStop}%
\bibitem [{\citenamefont {Brenner}\ \emph {et~al.}(2006)\citenamefont
  {Brenner}, \citenamefont {Cakirli},\ and\ \citenamefont
  {Casten}}]{Brenner2006PRC}%
  \BibitemOpen
  \bibfield  {author} {\bibinfo {author} {\bibfnamefont {D.~S.}\ \bibnamefont
  {Brenner}}, \bibinfo {author} {\bibfnamefont {R.~B.}\ \bibnamefont
  {Cakirli}}, \ and\ \bibinfo {author} {\bibfnamefont {R.~F.}\ \bibnamefont
  {Casten}},\ }\href {\doibase 10.1103/PhysRevC.73.034315} {\bibfield
  {journal} {\bibinfo  {journal} {Phys. Rev. C}\ }\textbf {\bibinfo {volume}
  {73}},\ \bibinfo {pages} {034315} (\bibinfo {year} {2006})}\BibitemShut
  {NoStop}%
\bibitem [{\citenamefont {Fu}\ \emph {et~al.}(2010)\citenamefont {Fu},
  \citenamefont {Jiang}, \citenamefont {Zhao},\ and\ \citenamefont
  {Arima}}]{Fu2010PRC}%
  \BibitemOpen
  \bibfield  {author} {\bibinfo {author} {\bibfnamefont {G.~J.}\ \bibnamefont
  {Fu}}, \bibinfo {author} {\bibfnamefont {H.}~\bibnamefont {Jiang}}, \bibinfo
  {author} {\bibfnamefont {Y.~M.}\ \bibnamefont {Zhao}}, \ and\ \bibinfo
  {author} {\bibfnamefont {A.}~\bibnamefont {Arima}},\ }\href {\doibase
  10.1103/PhysRevC.82.014307} {\bibfield  {journal} {\bibinfo  {journal} {Phys.
  Rev. C}\ }\textbf {\bibinfo {volume} {82}},\ \bibinfo {pages} {014307}
  (\bibinfo {year} {2010})}\BibitemShut {NoStop}%
\bibitem [{\citenamefont {Bender}\ and\ \citenamefont
  {Heenen}(2011)}]{Bender2011PRC}%
  \BibitemOpen
  \bibfield  {author} {\bibinfo {author} {\bibfnamefont {M.}~\bibnamefont
  {Bender}}\ and\ \bibinfo {author} {\bibfnamefont {P.-H.}\ \bibnamefont
  {Heenen}},\ }\href {\doibase 10.1103/PhysRevC.83.064319} {\bibfield
  {journal} {\bibinfo  {journal} {Phys. Rev. C}\ }\textbf {\bibinfo {volume}
  {83}},\ \bibinfo {pages} {064319} (\bibinfo {year} {2011})}\BibitemShut
  {NoStop}%
\bibitem [{\citenamefont {Wu}\ \emph {et~al.}(2016)\citenamefont {Wu},
  \citenamefont {Changizi},\ and\ \citenamefont {Qi}}]{Wu2016PRC}%
  \BibitemOpen
  \bibfield  {author} {\bibinfo {author} {\bibfnamefont {Z.}~\bibnamefont
  {Wu}}, \bibinfo {author} {\bibfnamefont {S.~A.}\ \bibnamefont {Changizi}}, \
  and\ \bibinfo {author} {\bibfnamefont {C.}~\bibnamefont {Qi}},\ }\href
  {\doibase 10.1103/PhysRevC.93.034334} {\bibfield  {journal} {\bibinfo
  {journal} {Phys. Rev. C}\ }\textbf {\bibinfo {volume} {93}},\ \bibinfo
  {pages} {034334} (\bibinfo {year} {2016})}\BibitemShut {NoStop}%
\bibitem [{\citenamefont {Zhang}\ and\ \citenamefont
  {Zhang}(2019)}]{ZhangW2019PRC}%
  \BibitemOpen
  \bibfield  {author} {\bibinfo {author} {\bibfnamefont {W.}~\bibnamefont
  {Zhang}}\ and\ \bibinfo {author} {\bibfnamefont {S.~Q.}\ \bibnamefont
  {Zhang}},\ }\href {\doibase 10.1103/PhysRevC.100.054303} {\bibfield
  {journal} {\bibinfo  {journal} {Phys. Rev. C}\ }\textbf {\bibinfo {volume}
  {100}},\ \bibinfo {pages} {054303} (\bibinfo {year} {2019})}\BibitemShut
  {NoStop}%
\bibitem [{\citenamefont {Wang}\ \emph {et~al.}(2024)\citenamefont {Wang},
  \citenamefont {Wang}, \citenamefont {Xu}, \citenamefont {Zhao},\ and\
  \citenamefont {Meng}}]{Wang2024PRL}%
  \BibitemOpen
  \bibfield  {author} {\bibinfo {author} {\bibfnamefont {Y.~P.}\ \bibnamefont
  {Wang}}, \bibinfo {author} {\bibfnamefont {Y.~K.}\ \bibnamefont {Wang}},
  \bibinfo {author} {\bibfnamefont {F.~F.}\ \bibnamefont {Xu}}, \bibinfo
  {author} {\bibfnamefont {P.~W.}\ \bibnamefont {Zhao}}, \ and\ \bibinfo
  {author} {\bibfnamefont {J.}~\bibnamefont {Meng}},\ }\href {\doibase
  10.1103/PhysRevLett.132.232501} {\bibfield  {journal} {\bibinfo  {journal}
  {Phys. Rev. Lett.}\ }\textbf {\bibinfo {volume} {132}},\ \bibinfo {pages}
  {232501} (\bibinfo {year} {2024})}\BibitemShut {NoStop}%
\bibitem [{\citenamefont {Cakirli}\ \emph {et~al.}(2025)\citenamefont
  {Cakirli}, \citenamefont {Blaum},\ and\ \citenamefont
  {Casten}}]{Cakirli2025FOP}%
  \BibitemOpen
  \bibfield  {author} {\bibinfo {author} {\bibfnamefont {R.~B.}\ \bibnamefont
  {Cakirli}}, \bibinfo {author} {\bibfnamefont {K.}~\bibnamefont {Blaum}}, \
  and\ \bibinfo {author} {\bibfnamefont {R.~F.}\ \bibnamefont {Casten}},\
  }\href {\doibase 10.3389/fphy.2025.1653635} {\bibfield  {journal} {\bibinfo
  {journal} {Front. Phys.}\ }\textbf {\bibinfo {volume} {13}},\ \bibinfo
  {pages} {1653635} (\bibinfo {year} {2025})}\BibitemShut {NoStop}%
\bibitem [{\citenamefont {Federman}\ and\ \citenamefont
  {Pittel}(1978)}]{Federman1978PLB}%
  \BibitemOpen
  \bibfield  {author} {\bibinfo {author} {\bibfnamefont {P.}~\bibnamefont
  {Federman}}\ and\ \bibinfo {author} {\bibfnamefont {S.}~\bibnamefont
  {Pittel}},\ }\href {\doibase https://doi.org/10.1016/0370-2693(78)90192-2}
  {\bibfield  {journal} {\bibinfo  {journal} {Phys. Lett. B}\ }\textbf
  {\bibinfo {volume} {77}},\ \bibinfo {pages} {29} (\bibinfo {year}
  {1978})}\BibitemShut {NoStop}%
\bibitem [{\citenamefont {Stoitsov}\ \emph {et~al.}(2007)\citenamefont
  {Stoitsov}, \citenamefont {Cakirli}, \citenamefont {Casten}, \citenamefont
  {Nazarewicz},\ and\ \citenamefont {Satu\l{}a}}]{Stoitsov2007PRL}%
  \BibitemOpen
  \bibfield  {author} {\bibinfo {author} {\bibfnamefont {M.}~\bibnamefont
  {Stoitsov}}, \bibinfo {author} {\bibfnamefont {R.~B.}\ \bibnamefont
  {Cakirli}}, \bibinfo {author} {\bibfnamefont {R.~F.}\ \bibnamefont {Casten}},
  \bibinfo {author} {\bibfnamefont {W.}~\bibnamefont {Nazarewicz}}, \ and\
  \bibinfo {author} {\bibfnamefont {W.}~\bibnamefont {Satu\l{}a}},\ }\href
  {\doibase 10.1103/PhysRevLett.98.132502} {\bibfield  {journal} {\bibinfo
  {journal} {Phys. Rev. Lett.}\ }\textbf {\bibinfo {volume} {98}},\ \bibinfo
  {pages} {132502} (\bibinfo {year} {2007})}\BibitemShut {NoStop}%
\bibitem [{\citenamefont {Ring}(1996)}]{Ring1996PPNP}%
  \BibitemOpen
  \bibfield  {author} {\bibinfo {author} {\bibfnamefont {P.}~\bibnamefont
  {Ring}},\ }\href {\doibase https://doi.org/10.1016/0146-6410(96)00054-3}
  {\bibfield  {journal} {\bibinfo  {journal} {Prog. Part. Nucl. Phys.}\
  }\textbf {\bibinfo {volume} {37}},\ \bibinfo {pages} {193} (\bibinfo {year}
  {1996})}\BibitemShut {NoStop}%
\bibitem [{\citenamefont {Ren}\ and\ \citenamefont {Zhao}(2020)}]{Ren2020PRC}%
  \BibitemOpen
  \bibfield  {author} {\bibinfo {author} {\bibfnamefont {Z.~X.}\ \bibnamefont
  {Ren}}\ and\ \bibinfo {author} {\bibfnamefont {P.~W.}\ \bibnamefont {Zhao}},\
  }\href {\doibase 10.1103/PhysRevC.102.021301} {\bibfield  {journal} {\bibinfo
   {journal} {Phys. Rev. C}\ }\textbf {\bibinfo {volume} {102}},\ \bibinfo
  {pages} {021301(R)} (\bibinfo {year} {2020})}\BibitemShut {NoStop}%
\bibitem [{\citenamefont {Ren}\ \emph {et~al.}(2019)\citenamefont {Ren},
  \citenamefont {Zhang}, \citenamefont {Zhao}, \citenamefont {Itagaki},
  \citenamefont {Maruhn},\ and\ \citenamefont {Meng}}]{Ren2019SCI}%
  \BibitemOpen
  \bibfield  {author} {\bibinfo {author} {\bibfnamefont {Z.~X.}\ \bibnamefont
  {Ren}}, \bibinfo {author} {\bibfnamefont {S.~Q.}\ \bibnamefont {Zhang}},
  \bibinfo {author} {\bibfnamefont {P.~W.}\ \bibnamefont {Zhao}}, \bibinfo
  {author} {\bibfnamefont {N.}~\bibnamefont {Itagaki}}, \bibinfo {author}
  {\bibfnamefont {J.~A.}\ \bibnamefont {Maruhn}}, \ and\ \bibinfo {author}
  {\bibfnamefont {J.}~\bibnamefont {Meng}},\ }\href {\doibase
  https://doi.org/10.1007/s11433-019-9412-3} {\bibfield  {journal} {\bibinfo
  {journal} {Sci. China-Phys. Mech. Astron.}\ }\textbf {\bibinfo {volume}
  {62}},\ \bibinfo {pages} {112062} (\bibinfo {year} {2019})}\BibitemShut
  {NoStop}%
\bibitem [{\citenamefont {Li}\ \emph {et~al.}(2020)\citenamefont {Li},
  \citenamefont {Ren},\ and\ \citenamefont {Zhao}}]{Li2020PRC}%
  \BibitemOpen
  \bibfield  {author} {\bibinfo {author} {\bibfnamefont {B.}~\bibnamefont
  {Li}}, \bibinfo {author} {\bibfnamefont {Z.~X.}\ \bibnamefont {Ren}}, \ and\
  \bibinfo {author} {\bibfnamefont {P.~W.}\ \bibnamefont {Zhao}},\ }\href
  {\doibase 10.1103/PhysRevC.102.044307} {\bibfield  {journal} {\bibinfo
  {journal} {Phys. Rev. C}\ }\textbf {\bibinfo {volume} {102}},\ \bibinfo
  {pages} {044307} (\bibinfo {year} {2020})}\BibitemShut {NoStop}%
\bibitem [{\citenamefont {Zhang}\ \emph {et~al.}(2022)\citenamefont {Zhang},
  \citenamefont {Ren}, \citenamefont {Zhao}, \citenamefont {Vretenar},
  \citenamefont {Nik\ifmmode \check{s}\else \v{s}\fi{}i\ifmmode~\acute{c}\else
  \'{c}\fi{}},\ and\ \citenamefont {Meng}}]{ZhangDD2022PRC}%
  \BibitemOpen
  \bibfield  {author} {\bibinfo {author} {\bibfnamefont {D.~D.}\ \bibnamefont
  {Zhang}}, \bibinfo {author} {\bibfnamefont {Z.~X.}\ \bibnamefont {Ren}},
  \bibinfo {author} {\bibfnamefont {P.~W.}\ \bibnamefont {Zhao}}, \bibinfo
  {author} {\bibfnamefont {D.}~\bibnamefont {Vretenar}}, \bibinfo {author}
  {\bibfnamefont {T.}~\bibnamefont {Nik\ifmmode \check{s}\else
  \v{s}\fi{}i\ifmmode~\acute{c}\else \'{c}\fi{}}}, \ and\ \bibinfo {author}
  {\bibfnamefont {J.}~\bibnamefont {Meng}},\ }\href {\doibase
  10.1103/PhysRevC.105.024322} {\bibfield  {journal} {\bibinfo  {journal}
  {Phys. Rev. C}\ }\textbf {\bibinfo {volume} {105}},\ \bibinfo {pages}
  {024322} (\bibinfo {year} {2022})}\BibitemShut {NoStop}%
\bibitem [{\citenamefont {Ren}\ \emph {et~al.}(2020)\citenamefont {Ren},
  \citenamefont {Zhao}, \citenamefont {Zhang},\ and\ \citenamefont
  {Meng}}]{Ren2020NPA}%
  \BibitemOpen
  \bibfield  {author} {\bibinfo {author} {\bibfnamefont {Z.~X.}\ \bibnamefont
  {Ren}}, \bibinfo {author} {\bibfnamefont {P.~W.}\ \bibnamefont {Zhao}},
  \bibinfo {author} {\bibfnamefont {S.~Q.}\ \bibnamefont {Zhang}}, \ and\
  \bibinfo {author} {\bibfnamefont {J.}~\bibnamefont {Meng}},\ }\href {\doibase
  https://doi.org/10.1016/j.nuclphysa.2020.121696} {\bibfield  {journal}
  {\bibinfo  {journal} {Nucl. Phys. A}\ }\textbf {\bibinfo {volume} {996}},\
  \bibinfo {pages} {121696} (\bibinfo {year} {2020})}\BibitemShut {NoStop}%
\bibitem [{\citenamefont {Yamagami}\ \emph {et~al.}(2001)\citenamefont
  {Yamagami}, \citenamefont {Matsuyanagi},\ and\ \citenamefont
  {Matsuo}}]{Yamagami2001NPA}%
  \BibitemOpen
  \bibfield  {author} {\bibinfo {author} {\bibfnamefont {M.}~\bibnamefont
  {Yamagami}}, \bibinfo {author} {\bibfnamefont {K.}~\bibnamefont
  {Matsuyanagi}}, \ and\ \bibinfo {author} {\bibfnamefont {M.}~\bibnamefont
  {Matsuo}},\ }\href {\doibase https://doi.org/10.1016/S0375-9474(01)00918-6}
  {\bibfield  {journal} {\bibinfo  {journal} {Nucl. Phys. A}\ }\textbf
  {\bibinfo {volume} {693}},\ \bibinfo {pages} {579} (\bibinfo {year}
  {2001})}\BibitemShut {NoStop}%
\bibitem [{\citenamefont {Meng}(2016)}]{Meng2016Book}%
  \BibitemOpen
  \bibinfo {editor} {\bibfnamefont {J.}~\bibnamefont {Meng}},\ ed.,\ \href@noop
  {} {\emph {\bibinfo {title} {{Relativistic Density Functional for Nuclear
  Structure, International Review of Nuclear Physics}}}},\ Vol.~\bibinfo
  {volume} {10}\ (\bibinfo  {publisher} {World Scientific, Singapore},\
  \bibinfo {year} {2016})\BibitemShut {NoStop}%
\bibitem [{\citenamefont {Vretenar}\ \emph {et~al.}(2005)\citenamefont
  {Vretenar}, \citenamefont {Afanasjev}, \citenamefont {Lalazissis},\ and\
  \citenamefont {Ring}}]{Vretenar2005PR}%
  \BibitemOpen
  \bibfield  {author} {\bibinfo {author} {\bibfnamefont {D.}~\bibnamefont
  {Vretenar}}, \bibinfo {author} {\bibfnamefont {A.~V.}\ \bibnamefont
  {Afanasjev}}, \bibinfo {author} {\bibfnamefont {G.~A.}\ \bibnamefont
  {Lalazissis}}, \ and\ \bibinfo {author} {\bibfnamefont {P.}~\bibnamefont
  {Ring}},\ }\href {\doibase https://doi.org/10.1016/j.physrep.2004.10.001}
  {\bibfield  {journal} {\bibinfo  {journal} {Phys. Rep.}\ }\textbf {\bibinfo
  {volume} {409}},\ \bibinfo {pages} {101} (\bibinfo {year}
  {2005})}\BibitemShut {NoStop}%
\bibitem [{\citenamefont {Nik\v{s}i\'{c}}\ \emph {et~al.}(2011)\citenamefont
  {Nik\v{s}i\'{c}}, \citenamefont {Vretenar},\ and\ \citenamefont
  {Ring}}]{Niksic2011PPNP}%
  \BibitemOpen
  \bibfield  {author} {\bibinfo {author} {\bibfnamefont {T.}~\bibnamefont
  {Nik\v{s}i\'{c}}}, \bibinfo {author} {\bibfnamefont {D.}~\bibnamefont
  {Vretenar}}, \ and\ \bibinfo {author} {\bibfnamefont {P.}~\bibnamefont
  {Ring}},\ }\href {\doibase https://doi.org/10.1016/j.ppnp.2011.01.055}
  {\bibfield  {journal} {\bibinfo  {journal} {Prog. Part. Nucl. Phys.}\
  }\textbf {\bibinfo {volume} {66}},\ \bibinfo {pages} {519} (\bibinfo {year}
  {2011})}\BibitemShut {NoStop}%
\bibitem [{\citenamefont {Ryssens}\ \emph {et~al.}(2015)\citenamefont
  {Ryssens}, \citenamefont {Hellemans}, \citenamefont {Bender},\ and\
  \citenamefont {Heenen}}]{Ryssens2015CPC}%
  \BibitemOpen
  \bibfield  {author} {\bibinfo {author} {\bibfnamefont {W.}~\bibnamefont
  {Ryssens}}, \bibinfo {author} {\bibfnamefont {V.}~\bibnamefont {Hellemans}},
  \bibinfo {author} {\bibfnamefont {M.}~\bibnamefont {Bender}}, \ and\ \bibinfo
  {author} {\bibfnamefont {P.-H.}\ \bibnamefont {Heenen}},\ }\href {\doibase
  https://doi.org/10.1016/j.cpc.2015.01.011} {\bibfield  {journal} {\bibinfo
  {journal} {Comput. Phys. Commun.}\ }\textbf {\bibinfo {volume} {190}},\
  \bibinfo {pages} {231} (\bibinfo {year} {2015})}\BibitemShut {NoStop}%
\bibitem [{\citenamefont {Zhao}\ \emph {et~al.}(2010)\citenamefont {Zhao},
  \citenamefont {Li}, \citenamefont {Yao},\ and\ \citenamefont
  {Meng}}]{Zhao2010PRC}%
  \BibitemOpen
  \bibfield  {author} {\bibinfo {author} {\bibfnamefont {P.~W.}\ \bibnamefont
  {Zhao}}, \bibinfo {author} {\bibfnamefont {Z.~P.}\ \bibnamefont {Li}},
  \bibinfo {author} {\bibfnamefont {J.~M.}\ \bibnamefont {Yao}}, \ and\
  \bibinfo {author} {\bibfnamefont {J.}~\bibnamefont {Meng}},\ }\href {\doibase
  10.1103/PhysRevC.82.054319} {\bibfield  {journal} {\bibinfo  {journal} {Phys.
  Rev. C}\ }\textbf {\bibinfo {volume} {82}},\ \bibinfo {pages} {054319}
  (\bibinfo {year} {2010})}\BibitemShut {NoStop}%
\bibitem [{\citenamefont {Wang}\ \emph {et~al.}(2021)\citenamefont {Wang},
  \citenamefont {Huang}, \citenamefont {Kondev}, \citenamefont {Audi},\ and\
  \citenamefont {Naimi}}]{Wang2021CPC}%
  \BibitemOpen
  \bibfield  {author} {\bibinfo {author} {\bibfnamefont {M.}~\bibnamefont
  {Wang}}, \bibinfo {author} {\bibfnamefont {W.~J.}\ \bibnamefont {Huang}},
  \bibinfo {author} {\bibfnamefont {F.~G.}\ \bibnamefont {Kondev}}, \bibinfo
  {author} {\bibfnamefont {G.}~\bibnamefont {Audi}}, \ and\ \bibinfo {author}
  {\bibfnamefont {S.}~\bibnamefont {Naimi}},\ }\href {\doibase
  10.1088/1674-1137/abddaf} {\bibfield  {journal} {\bibinfo  {journal} {Chin.
  Phys. C}\ }\textbf {\bibinfo {volume} {45}},\ \bibinfo {pages} {030003}
  (\bibinfo {year} {2021})}\BibitemShut {NoStop}%
\bibitem [{\citenamefont {Hamaker}\ \emph {et~al.}(2021)\citenamefont
  {Hamaker}, \citenamefont {Leistenschneider}, \citenamefont {Jain},
  \citenamefont {Bollen}, \citenamefont {Giuliani}, \citenamefont {Lund},
  \citenamefont {Nazarewicz}, \citenamefont {Neufcourt}, \citenamefont
  {Nicoloff}, \citenamefont {Puentes}, \citenamefont {Ringle}, \citenamefont
  {Sumithrarachchi},\ and\ \citenamefont {Yandow}}]{Hamaker2021Nature}%
  \BibitemOpen
  \bibfield  {author} {\bibinfo {author} {\bibfnamefont {A.}~\bibnamefont
  {Hamaker}}, \bibinfo {author} {\bibfnamefont {E.}~\bibnamefont
  {Leistenschneider}}, \bibinfo {author} {\bibfnamefont {R.}~\bibnamefont
  {Jain}}, \bibinfo {author} {\bibfnamefont {G.}~\bibnamefont {Bollen}},
  \bibinfo {author} {\bibfnamefont {S.~A.}\ \bibnamefont {Giuliani}}, \bibinfo
  {author} {\bibfnamefont {K.}~\bibnamefont {Lund}}, \bibinfo {author}
  {\bibfnamefont {W.}~\bibnamefont {Nazarewicz}}, \bibinfo {author}
  {\bibfnamefont {L.}~\bibnamefont {Neufcourt}}, \bibinfo {author}
  {\bibfnamefont {C.~R.}\ \bibnamefont {Nicoloff}}, \bibinfo {author}
  {\bibfnamefont {D.}~\bibnamefont {Puentes}}, \bibinfo {author} {\bibfnamefont
  {R.}~\bibnamefont {Ringle}}, \bibinfo {author} {\bibfnamefont {C.~S.}\
  \bibnamefont {Sumithrarachchi}}, \ and\ \bibinfo {author} {\bibfnamefont
  {I.~T.}\ \bibnamefont {Yandow}},\ }\href {\doibase
  https://doi.org/10.1038/s41567-021-01395-w} {\bibfield  {journal} {\bibinfo
  {journal} {Nat. Phys.}\ }\textbf {\bibinfo {volume} {17}},\ \bibinfo {pages}
  {1408} (\bibinfo {year} {2021})}\BibitemShut {NoStop}%
\bibitem [{\citenamefont {Erler}\ \emph {et~al.}(2012)\citenamefont {Erler},
  \citenamefont {Birge}, \citenamefont {Kortelainen}, \citenamefont
  {Nazarewicz}, \citenamefont {Olsen}, \citenamefont {Perhac},\ and\
  \citenamefont {Stoitsov}}]{Erler2012Nature}%
  \BibitemOpen
  \bibfield  {author} {\bibinfo {author} {\bibfnamefont {J.}~\bibnamefont
  {Erler}}, \bibinfo {author} {\bibfnamefont {N.}~\bibnamefont {Birge}},
  \bibinfo {author} {\bibfnamefont {M.}~\bibnamefont {Kortelainen}}, \bibinfo
  {author} {\bibfnamefont {W.}~\bibnamefont {Nazarewicz}}, \bibinfo {author}
  {\bibfnamefont {E.}~\bibnamefont {Olsen}}, \bibinfo {author} {\bibfnamefont
  {A.~M.}\ \bibnamefont {Perhac}}, \ and\ \bibinfo {author} {\bibfnamefont
  {M.}~\bibnamefont {Stoitsov}},\ }\href {\doibase
  https://doi.org/10.1038/nature11188} {\bibfield  {journal} {\bibinfo
  {journal} {Nature}\ }\textbf {\bibinfo {volume} {486}},\ \bibinfo {pages}
  {509} (\bibinfo {year} {2012})}\BibitemShut {NoStop}%
\bibitem [{\citenamefont {Agbemava}\ \emph {et~al.}(2014)\citenamefont
  {Agbemava}, \citenamefont {Afanasjev}, \citenamefont {Ray},\ and\
  \citenamefont {Ring}}]{Agbemava2014PRC}%
  \BibitemOpen
  \bibfield  {author} {\bibinfo {author} {\bibfnamefont {S.~E.}\ \bibnamefont
  {Agbemava}}, \bibinfo {author} {\bibfnamefont {A.~V.}\ \bibnamefont
  {Afanasjev}}, \bibinfo {author} {\bibfnamefont {D.}~\bibnamefont {Ray}}, \
  and\ \bibinfo {author} {\bibfnamefont {P.}~\bibnamefont {Ring}},\ }\href
  {\doibase 10.1103/PhysRevC.89.054320} {\bibfield  {journal} {\bibinfo
  {journal} {Phys. Rev. C}\ }\textbf {\bibinfo {volume} {89}},\ \bibinfo
  {pages} {054320} (\bibinfo {year} {2014})}\BibitemShut {NoStop}%
\bibitem [{\citenamefont {Wu}\ and\ \citenamefont {Zhao}(2020)}]{Wu2020PRC}%
  \BibitemOpen
  \bibfield  {author} {\bibinfo {author} {\bibfnamefont {X.~H.}\ \bibnamefont
  {Wu}}\ and\ \bibinfo {author} {\bibfnamefont {P.~W.}\ \bibnamefont {Zhao}},\
  }\href {\doibase 10.1103/PhysRevC.101.051301} {\bibfield  {journal} {\bibinfo
   {journal} {Phys. Rev. C}\ }\textbf {\bibinfo {volume} {101}},\ \bibinfo
  {pages} {051301} (\bibinfo {year} {2020})}\BibitemShut {NoStop}%
\bibitem [{\citenamefont {Möller}\ \emph {et~al.}(2016)\citenamefont
  {Möller}, \citenamefont {Sierk}, \citenamefont {Ichikawa},\ and\
  \citenamefont {Sagawa}}]{Moller2016ADNDT}%
  \BibitemOpen
  \bibfield  {author} {\bibinfo {author} {\bibfnamefont {P.}~\bibnamefont
  {Möller}}, \bibinfo {author} {\bibfnamefont {A.}~\bibnamefont {Sierk}},
  \bibinfo {author} {\bibfnamefont {T.}~\bibnamefont {Ichikawa}}, \ and\
  \bibinfo {author} {\bibfnamefont {H.}~\bibnamefont {Sagawa}},\ }\href
  {\doibase https://doi.org/10.1016/j.adt.2015.10.002} {\bibfield  {journal}
  {\bibinfo  {journal} {Atomic Data and Nuclear Data Tables}\ }\textbf
  {\bibinfo {volume} {109-110}},\ \bibinfo {pages} {1} (\bibinfo {year}
  {2016})}\BibitemShut {NoStop}%
\end{thebibliography}


\end{document}